\documentclass[manuscript,screen,review]{acmart}

\usepackage{longtable}
\usepackage{array}
\usepackage{wrapfig}
\usepackage{xcolor}

\AtBeginDocument{%
  \providecommand\BibTeX{{%
    \normalfont B\kern-0.5em{\scshape i\kern-0.25em b}\kern-0.8em\TeX}}}

\setcopyright{acmcopyright}
\copyrightyear{2021}
\acmYear{2021}

\begin{document}

\title{Survey of Information Encoding Techniques for DNA}

\author{Thomas Heinis}
\email{t.heinis@imperial.ac.uk}
\author{Roman Sokolovskii}
\email{r.sokolovskii@imperial.ac.uk}
\author{Jamie J. Alnasir}
\email{j.alnasir@imperial.ac.uk}
\affiliation{%
  \institution{Imperial College London}
  \streetaddress{180 Queen's Gate, South Kensington}
  \city{London}
  \state{England}
  \country{UK}
  \postcode{SW7 2AZ}
}


\renewcommand{\shortauthors}{Heinis \textit{et al.}}


\begin{abstract}
The yearly global production of data is growing exponentially, outpacing the capacity of existing storage media, such as tape and disk, and surpassing our ability to store it. DNA storage---the representation of arbitrary information as sequences of nucleotides---offers a promising storage medium. DNA is nature's information-storage molecule of choice and has a number of key properties: it is extremely dense, offering the theoretical possibility of storing 455 EB/g; it is durable, with a half-life of approximately 520 years that can be increased to thousands of years when DNA is chilled and stored dry; and it is amenable to automated synthesis and sequencing. Furthermore, biochemical processes that act on DNA potentially enable highly parallel data manipulation.

Whilst biological information is encoded in DNA via a specific mapping from triplet sequences of nucleotides to amino acids, DNA storage is not limited to a single encoding scheme, and there are many possible ways to map data to chemical sequences of nucleotides for synthesis, storage, retrieval and data manipulation. However, there are several biological, error-tolerance and information-retrieval considerations that an encoding scheme needs to address to be viable.

This comprehensive review focuses on comparing existing work done in encoding arbitrary data within DNA in terms of their encoding schemes, methods to address biological constraints and measures to provide error correction. We compare encoding approaches on the overall information density and coverage they achieve, as well as the data-retrieval method they use (i.e., sequential or random access). We also discuss the background and evolution of the encoding schemes.
\end{abstract}



\maketitle

\section{Introduction}

DNA storage, whereby DNA is used as a medium to store arbitrary data, has been proposed as a method to address the storage crisis that is currently arising---the world's data production is outpacing our ability to store it~\cite{cox2001long, Goldman2013TowardsDNA, Grass2015RobustCodes}. Furthermore, rapid advancements in information-storage technologies mean that, as a result of hardware and software obsolescence as well as physical degradation, data currently stored in magnetic or optical media will likely be unrecoverable in a century or less~\cite{Bornholt2016ASystem}. Hence, new storage methods are required to allow data to be retrieved after longer periods~\cite{Church2012Next-generationDNA}.

As a medium for immutable, high-latency archival storage, DNA offers several important advantages: It is extremely durable---without treating it to improve durability, DNA has a half-life of approximately 520 years~\cite{Allentoft2012TheFossils}. It is exceptionally dense---theoretically, it is possible to achieve an information density of 455 EB/g, i.e., to store 455 million Terabytes in one gram of DNA~\cite{Church2012Next-generationDNA}. Besides, as a result of DNA being the information-storage molecule of all known organisms, all materials for synthesising DNA are abundantly available.



DNA storage uses an encoding scheme to map arbitrary information to a set of DNA sequences. A DNA sequence is made of nucleotides that each contains one of the four distinct bases: adenine (A), thymine (T), guanine (G) and cytosine (C). The resulting mapped sequences are then physically created as DNA molecules (often referred to as oligonucleotides, which means short DNA sequences) through enzymatic or chemical synthesis. The synthesised DNA molecules can be single-stranded (ssDNA) or double-stranded (dsDNA), i.e., with two DNA strands forming a double helix where each base in one strand is connected to its complement in the other strand according to the pairing of A with T and C with G. The synthesised oligonucleotides are stored for extended periods. 

\begin{figure}[h]
  \begin{center}
    \includegraphics[width=.95\textwidth]{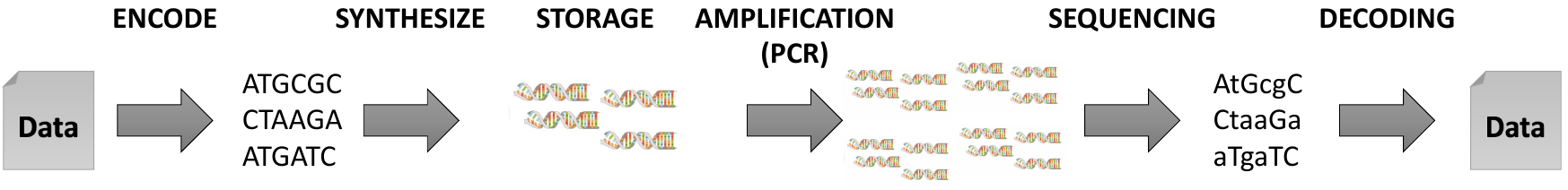}
    \caption{Illustration of the the end-to-end DNA storage workflow from writing via DNA synthesis to reading via DNA sequencing.}
    \label{fig:dnadatastroage}
  \end{center}
\end{figure}

To read the information back, DNA sequencing is used. Two primary types of sequencing have evolved to date, namely sequencing by synthesis (SBS) and nanopore sequencing. SBS has a limited read-length of hundreds to thousands of nucleotides (nt) and requires significant molecular biological preparatory steps, including fragmentation (to the sizes tolerated by the sequencer) and amplification via polymerase chain reaction (PCR). PCR copies DNA sequences hundreds to millions of times, thus ensuring enough DNA material is available for sequencing. Nanopore sequencers, on the other hand, do not strictly require fragmentation or amplification as they are capable of sequencing long DNA molecules. In both cases, copies of the same sequence are typically read multiple times (sequencing 5 copies of the same sequence would be referred to as sequencing with a coverage of 5). The reads of the same sequence are used to obtain a consensus on each nucleotide at each position. Then the consensus sequences are decoded, i.e., mapped back to their binary representation, and error correction is applied where possible and needed to retrieve the stored data with high fidelity. Figure~\ref{fig:dnadatastroage} illustrates the end-to-end process.

\section{Comparing DNA storage approaches}

The central step of the DNA storage pipeline is the encoding of information into a sequence of nucleotides to be synthesised. In this survey, we study different approaches to this encoding. Although some clearly have only been designed for artistic purposes and not for information storage, we still include them for completeness. We compare approaches with respect to the storage mechanism, information density, required coverage, error detection and correction mechanisms, biological constraints considered, access mechanism and the type and size of the stored data.

We first introduce the criteria for comparison and then discuss the encoding approaches, grouping them into \textit{initial} and \textit{advanced} approaches (Section~\ref{sec:initial} and~\ref{sec:advanced}, respectively).
This grouping roughly corresponds to the chronological order and reflects their increasing level of sophistication.
We remark that this survey focuses specifically on the encoding step of the DNA storage pipeline (see Fig.~\ref{fig:dnadatastroage}). For a more general review of DNA storage, we refer the reader to~\cite{ceze2019molecular,dong2020dna}. Compared with these reviews, our survey describes in more detail the specific encoding schemes and their properties, notably with respect to relevant biological constraints.

\subsection{Criteria for Comparison}\label{lblCriteria}

The viability of an approach to DNA storage depends on a number of factors. These form the criteria of our comparison and are as follows:

\begin{itemize}
\item \textbf{Storage medium}: there are several options for storing DNA. It is possible to store information \textit{in vivo} by inserting the DNA into the genetic information of a living organism, e.g., the bacteria \textit{E. coli, B. subtilis S. cerevisiae} or \textit{D. radiodurans}. However, errors may be introduced when the cells divide (during DNA recombination), and 
the amount of stored information is limited by how much extraneous DNA the host organism can tolerate without being incapacitated. DNA can also be stored \textit{in vitro}, e.g., by hybridisation to beads, or by storing within wells or microplates.

\item \textbf{Biological constraints}: the encoding of data into DNA must adhere to multiple design restrictions. Not all possible sequences of nucleotides can be synthesised. For example, long homopolymers, i.e., long sequences of the same nucleotide, are especially difficult to synthesise. Similarly, extremes in GC content, i.e., extremes in the prevalence of G or C in the sequence, should also be avoided as GC-rich or GC-poor sequences pose problems for both synthesis and sequencing \cite{dohm2008substantial}.

\item \textbf{Error correction}: both synthesis and sequencing are inherently prone to errors in the form of substitutions, deletions and insertions of nucleotides. Strand/sequence breakage may also occur in storing the DNA. An encoding must therefore be designed to be resilient to errors through error correction codes or replication (i.e., through storing the same sequence multiple times).

\item \textbf{Information density}: synthesis of DNA using current technologies is expensive, hence an encoding scheme must map as many bits as possible to one nucleotide, and a crucial metric---information density---assesses how much information (bits) can be stored per nucleotide. Several factors affect information density. For example, if more nucleotides in a sequence are used to store error detection and correction information, there are fewer nucleotides available for encoding the payload (i.e., the actual data to be stored), and information density decreases. Information density therefore strongly depends on the features of the encoding and cannot be considered in isolation. We provide a detailed description of how we calculate information density in Section~\ref{sse:density} below. 

\item \textbf{Coverage}: in principle, to minimise the cost of synthesis, the system designer needs to aim for the highest possible information density. However, the higher the information density, the less robust the system will be in the presence of errors. To correct the errors, it will be necessary to read more copies of the data; in other words, it will be necessary to increase coverage. There exist a fundamental trade-off between information density and coverage, and encoding methods should be compared on both counts. Although differences in experimental tools and setups make direct comparison difficult, we relay the minimum coverage that was sufficient for successful recovery whenever it is reported, which is typically the case for more recent approaches. (Some works do provide enough information to calculate coverage but do not attempt a sub-sampling experiment to determine the minimum sufficient coverage; we mark those cases explicitly in summary Table~\ref{table:comparison} at the end of the paper.)

\item \textbf{Data-retrieval method}: there are two general methods for recovering stored data that are applicable to DNA storage: (i) Sequential access, where the entire DNA needs to be sequenced to reconstruct the information even if only small parts of it are requested. Sequential access can be a slow and costly process. (ii) Random access, where subsets of the data can be selectively read without having to sequence all information. Random access is most often implemented via selective PCR amplification. This requires the placement of primers (i.e., additional nucleotide subsequences) at the 5' (head) and 3' (tail) ends of the DNA oligonucleotides, which lowers information density.\footnote{PCR amplification is known to be prone to bias and significant dropout rates, especially when performed over DNA pools with small numbers of physical copies of each oligonucleotide~\cite{chen2020quantifying}. Some alternatives to PCR amplification for random access in DNA storage have been proposed: In~\cite{lin2020dynamic}, Lin et al. use dsDNA fragments with overhanging ssDNA primers attached to them; a subset of strands that shares the same primer can be selectively fished out of the DNA pool with magnetic beads attached to the strand that is complementary to the primer.
Another work~\cite{banal2021random} achieves random access by encapsulating data-carrying strands and attaching ssDNA barcodes to the capsules. The barcodes represent the encapsulated data in such a way so as to allow both querying the database for specific files and performing Boolean metadata-based search and retrieval using fluorescent tags and fluorescence-activated sorting.}

\end{itemize}

\subsection{Information Density}\label{sse:density}

In order to appropriately review the encoding schemes, we calculate the overall information density in \textit{bits per nucleotide} (b/nt), which includes the encoded payload together with the additional subsequences such as primers needed for reading, amplification, etc. We also calculated the information density for the payload only and include both values in Table \ref{table:comparison}. Information density is calculated as
\begin{equation}\label{eqn:infoden}
 {\text \\bits\_per\_nt} = \frac{num{\text \_}symbols \times {bits{\text \_}per{\text \_}symbol} }{total{\text \_}nucleotides}\,.
\end{equation}

The difference between the overall and payload-only information density is in the term $total{\text\_}nucleotides$: for the overall information density, $total{\text\_}nucleotides$ includes the primer and index subsequences, whereas for the \textit{payload-only} information density the number of nucleotides used for optional, additional information such as primers and indexes is deducted from $total{\text\_}nucleotides$ in~\eqref{eqn:infoden}.
We remark that we count all payload-dependent redundancy in $total{\text\_}nucleotides$ for the payload-only information density. This includes parity-check strands or subsequences, cyclic redundancy checks and the lengthening of the strands due to constrained coding for biological constraints: we regard the redundancy thus introduced to correct, detect or prevent errors as an integral part of the payload.

Some approaches compress the data as a part of their encoding process. This is especially common among earlier approaches that deal with textual information. In those cases, the effect of compression inflates information density. We mark the information density figures that are affected by compression in Table~\ref{table:comparison} to highlight that the approaches that do use compression should not be compared with those that do not.

We mainly rely on the overall information density for comparison, but in some cases we also mention the payload-only information density, particularly when there is a large difference between the two due to the use of additional nucleotides for primers and indexes.

\section{Initial Approaches}\label{sec:initial}
\subsection{Microvenus~\texorpdfstring{\cite{Davis1996Microvenus}}{}}\label{sse:microvenus}
The first use of DNA as a storage device had primarily artistic purposes and was not meant to provide read and write functionality~\cite{Davis1996Microvenus}. The idea was to incorporate artificial (non-biological) information into DNA, and more specifically, into living bacteria (\textit{E. coli}). The information encoded is the ancient Germanic rune for \textit{life} and the \textit{female earth}, bit-mapped to a binary representation. This resulted in a 5$\times$7 or 7$\times$5 figure, totaling 35 bits. To translate the bits to DNA nucleotides, a phase-change model is used. In this model, each nucleotide indicates how may times a bit (0 or 1) is repeated before a change to another bit occurs:

\begin{center}
    $X \rightarrow C$ \quad $XX \rightarrow T$ \quad $XXX \rightarrow A$ \quad $XXXX \rightarrow G$
\end{center}  

With this code, the string \texttt{10101} is encoded as \texttt{CCCCC}, as each binary digit occurs once before changing, and \texttt{0111000} becomes \texttt{CAA}. The entire 35-bit message, which in this case starts with a 1, is therefore encoded into a sequence of 18 nucleotides, namely \texttt{CCCCCCAACGCGCGCGCT}. An additional sequence \texttt{CTTAAAGGGG} is prepended to indicate the start of the encoded message, and hints at the nature of the encoding nucleotides which represent repeating binary bits. This produces a DNA sequence with a final length of 28 nucleotides which was then implanted into a living cell, \textit{E. coli}, using the recombination molecular biology technique.

The information density of this approach depends on the exact message to be encoded. If, for example, the same bit is repeated multiple times, this approach only requires very few nucleotides and essentially compresses the information. The example message, using their encoding, allows for an overall information density of 1.25 b/nt (35 bits are encoded in 28 nucleotides). If the payload alone, excluding the prepended 10-nt sequence, is encoded using the same scheme, the information density is 1.94 b/nt.

One disadvantage of the encoding approach is the absence of any type of error detection or correction. Another is that it does not take into account biological constraints on the resulting DNA sequence such as long homopolymers or adequate GC content, making synthesis and sequencing challenging. To retrieve the information, all data has to be sequenced, thus only allowing sequential access to it. Furthermore, a specific downside of this approach is that DNA sequences are not uniquely decodable. As the nucleotides encode \textit{changes}, we can translate, for example, a \texttt{C} as either a \texttt{0} or \texttt{1}, and this generalises to any sequence: e.g., \texttt{TCA} can be decoded as either \texttt{001000} or \texttt{110111}.

\subsection{Genesis~\texorpdfstring{\cite{Kac1999GENESIS}}{}}\label{sse:genesis}
Similarly, to be understood as primarily art, is the Genesis project~\cite{Kac1999GENESIS}, which has the goal of encoding a sentence from \textit{The Book of Genesis}. The text used is \textit{Let man have dominion over the fish of the sea, and over the fowl of the air, and over every living thing that moves upon the earth}. The text is translated to Morse Code which is then again translated to a nucleotide sequence. Both the text and the Morse Code fit the symbolism of "genesis" (the Morse Code being one of the first technologies of the information age). 

The Morse code is translated to nucleotides as follows:

\begin{center}
    $Dash \rightarrow T$ \quad $Dot \rightarrow C$ \quad $Space \ (Word) \rightarrow A$ \quad $Space \ (Letter) \rightarrow G$
\end{center}  

Furthermore, for this encoding, the length of the DNA sequence depends on the exact message encoded as different letters have Morse code representations that vary in length. For the message used, 427 nucleotides are needed (using ITU Morse code). Assuming 5 bits are sufficient to encode a character (only adequate for characters), this results in an information density of 1.52 b/nt (130 chars. x 5 bits/char. / 427 nt), i.e., somewhat more than the previous approach.

The resulting encoding addresses one of the issues of the previous encoding, i.e., it encodes information without ambiguity. Still, the resulting DNA sequences may suffer from problems like homopolymer runs and imbalanced GC content and can only be read sequentially. Also, due to mutations in the bacteria, it was not possible to recover the original message (undesired modifications occurred) as no error detection or correction means were used in the encoding.

\subsection{Long-Term Storage of Information in DNA~\texorpdfstring{\cite{Bancroft2001Long-TermDNA}}{}}\label{sse:bancroft}
The first approach with the goal of storing and also reading information~\cite{Bancroft2001Long-TermDNA} splits the entire data in chuncks to be stored in information DNA (iDNA) sequences while the metadata (e.g., the order of the chunks) is stored in what are called polyprimer key (PPK) sequences.

More specifically, the iDNA contains a data chunk in its information segment which is flanked by a common forward (F) and reverse (R) primer (approximately 10-20 nucleotides) needed for PCR amplification. The iDNA further contains a sequencing primer (similar length), which is essentially an identifier for the iDNA sequence, and a common spacer (3-4 nucleotides) that indicates the start of the information segment. 

\begin{figure}[h]
    \centering
    \includegraphics[width=0.9\linewidth]{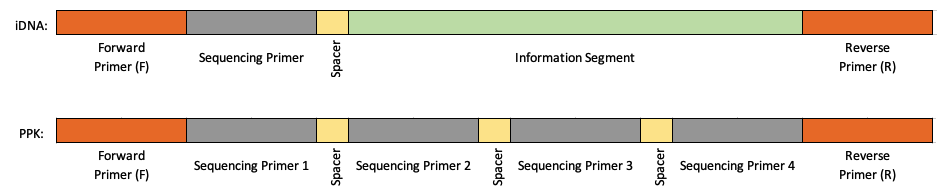}
    \caption{Illustrations of the structure of the iDNA and PPK sequences. The iDNA carries the information whereas the PPK stores the order of the information segments in the iDNA sequences.}
    \label{fig:bancroft_strands}
\end{figure}

The PPK sequence---the approach used only one such sequence---is flanked by the same forward and reverse primers and, crucially, contains the unique primers identifying each iDNA and the order in which they have to be read. Both types of sequences are illustrated in Figure~\ref{fig:bancroft_strands}.

For the retrieval process, the PPK is first amplified using PCR and sequenced to establish the order in which the iDNA sequences have to be read to reconstruct the initial data. All iDNA sequences are then also amplified and sequenced.

For the experiment, the text \textit{It was the best of times it was the worst of times} and \textit{It was the age of foolishness it was the epoch of belief} from \textit{A Tale of Two Cities} by Charles Dickens was chosen. The sentences were picked on the basis that some words were repeated multiple times and thus acted as a test for the robustness of the encoding scheme, as repeating data can lead to undesirable effects, such as, for example, homopolymer runs or extremes in GC content.

The text is encoded only using the three nucleotides \texttt{A}, \texttt{C} and \texttt{T}, while the sequencing primers are created using all 4 nucleotides, with the additional constraint that each fourth position is a \texttt{G}. Given this encoding approach, information segments cannot be mistaken for sequencing primers as only the latter contain \texttt{G} nucleotides.

The text is processed alphabetically by a ternary code that starts with \texttt{A} and alternates \texttt{C} and \texttt{T} in the third, second and first positions, e.g., the first letters of the alphabet are encoded as:

\begin{center}
    $A \rightarrow 000 \rightarrow \texttt{AAA}$ \quad
    $B \rightarrow 001 \rightarrow \texttt{AAC}$ \quad
    $C \rightarrow 002 \rightarrow \texttt{AAT}$ \quad
    $...$
\end{center}  

The two sentences were encoded and recovered successfully. As there are only 27 combinations of the three nucleotides, encoding more than 27 different letters would scale less efficiently as multiple triplets would be required for each letter. Further, using only one PPK inherently limits the amount of data that can be stored. However, the authors argue that one PPK should be enough to store the metadata (e.g., the order) for all iDNA sequences in a microplate.

The method of encoding the information does not avoid homopolymers of length three or more. A further issue is GC content: not using the \texttt{G} nucleotide in the information segment means that encoded sequences may exhibit an extreme GC content. Neither does the approach use error detection or correction codes.

Although not specifically designed for it, the approach allows for random access to the data. The PPK can be read out to retrieve the primers needed to read specific iDNA sequences.

iDNA sequences do not strictly have a limit in length, but the state of commercial synthesis technologies at the time the work was carried out meant two of their iDNA sequences of 232 and 247 nt were too long to be synthesised and were therefore constructed using overlapping oligonucleotides. The forward and reverse primers require a total of 40 nucleotides, the sequencing primer requires a further 20 nucleotides and the spacer ideally only requires 4 nucleotides (as opposed to the experiments where 19 nucleotides were used for the spacer), leaving 136 nucleotides for encoding information which amounts to about 45 letters given the encoding used. Assuming 5 bits are used to encode a letter, this results in an information density---for the payload only---of 1.65 b/nt. 

For their experiments, two iDNA sequences were used (one for each sentence encoded), with lengths of 232 and 247 nucleotides, respectively. For the sake of analysing the information density, we assume iDNA sequences of 232 and 247 nucleotides. The PPK contains forward and reverse primers of 20 nucleotides each, the two sequencing primers of each iDNA sequence each of 20 nucleotides length and two spacers with 4 nucleotides each, totaling 88 nucleotides. Factoring in the length of the PPK, this approach encodes to a final, overall information density of 0.94 b/nt. 

\subsection{Organic Data Memory Using the DNA Approach~\texorpdfstring{\cite{Wong2003OrganicApproach}}{}}\label{sse:wong}
A further approach~\cite{Wong2003OrganicApproach} studied encoding and storage in extreme conditions. The objective of this project is the development of a solution that can survive under extreme conditions, such as ultraviolet, partial vacuum, oxidation, high temperatures and radiation. For this purpose, the information was stored in the DNA of two well-studied bacteria, \textit{E. coli} and \textit{Deinococcus radiodurans}. The latter is particularly well-suited for extreme environments.

A similar encoding to that in \cite{Bancroft2001Long-TermDNA} (Section~\ref{sse:bancroft}) is used, but all four nucleotides are used to form the triplets. With this encoding, 64 symbols can be encoded, including letters, digits and punctuation. The first mappings between binary representation and nucleotides appear to be arbitrarily picked, i.e., to cycle through the nucleotides, and are as follows:

\begin{center}
    $0 \rightarrow \texttt{AAA}$ \quad
    $1 \rightarrow \texttt{AAC}$ \quad
    $2 \rightarrow \texttt{AAG}$ \quad
    $3 \rightarrow \texttt{AAT}$ \quad
    $4 \rightarrow \texttt{ACA}$ \quad
    $...$
\end{center}

\begin{wrapfigure}{L}{0.3\textwidth}
\centering
\includegraphics[width=0.25\textwidth]{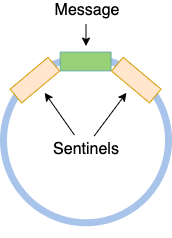}
\caption{Illustration of the sentinels flanking the message in the plasmid.}
\label{fig:flanking}
\end{wrapfigure}

To avoid interfering with the natural processes of the organism and to find the encoded message in the genome of the host organism, it is flanked by two sentinels, essentially two DNA sequences (see illustration in Figure~\ref{fig:flanking}). These two sequences must satisfy two constraints. First, the sentinels must not occur naturally in the host organism such that no sentinel will be mistaken for natural DNA in the host. Second, the sentinels need to contain triplets \texttt{TAG} or \texttt{TAA} that act as markers which tell the bacterium to stop translating the sequence (to proteins) so as not to interfere with natural processes. Analysing the entire genomes of \textit{E. coli} and \textit{Deinococcus radiodurans}, the authors found 25 such sequences with a length of 20 nucleotides.

For the experiment, fragments from the song "It's a Small World" were used, and the authors successfully stored and retrieved seven artificially synthesized fragments of length between 57 to 99 nucleotides, each flanked by two sentinels picked from the set of 25 sentinels, in seven individual bacteria. Since the flanking sequences are known, they can be used as primers and the retrieval is then performed using PCR and subsequent sequencing.

As discussed for the encoding in~\cite{Bancroft2001Long-TermDNA} (Section~\ref{sse:bancroft}), encoding symbols as triplets limits the alphabet and hence the amount of information that can be encoded---a maximum of 64 symbols can be used which, in most cases, is too restrictive to be of practical use unless multiple triplets are used which then decreases the information density that can be achieved. Another scalability issue is that the genome of the host has to be sequenced and then searched for appropriate locations to insert messages as well as sentinels. Even for microorganisms the search space is in the realm of billions, making the task computationally expensive. Finally, storing information in living organisms makes it exposed to mutations and detrimental environmental factors. To alleviate these issues, one of the ``toughest'' microorganisms known was used (\textit{Deinococcus}), but generally the problem of mutations remains unresolved, and the bacterium would likely not be able to accommodate large amounts of information. Care must also be taken not to damage the bacterium by it accidentally translating or transcribing DNA that is meant for information storage and is not part of its genome. Finally, homopolymers and GC content are not considered in the encoding. 

In their experiments, the authors encode 182 characters into 826 nucleotides (including the sentinels). Assuming 7 bits are needed to encode one letter/character (given that the encoding scheme allows for an alphabet of up to 128 letters), this yields an overall information density of 1.54 b/nt (182 chars. x 7 bits / 826 nt).

\subsection{Hiding Messages in DNA Microdots~\texorpdfstring{\cite{microdot}}{}}\label{sse:microdot}
A similar encoding approach~\cite{microdot} is used to encode messages and hiding them in a microdot. Alphanumeric values are mapped to sequences of three nucleotides:

\begin{center}
    $A \rightarrow \texttt{CGA}$ \quad
    $B \rightarrow \texttt{CCA}$ \quad
    $C \rightarrow \texttt{GTT}$ \quad
    $D \rightarrow \texttt{TTG}$ \quad
    $E \rightarrow \texttt{GGT}$ \quad
    $...$
\end{center}  

The mapping also includes special characters in a total of 36 symbols. A total of 64 symbols could be encoded with three nucleotides and so 28 sequences are not used. By picking the sequences carefully, one may avoid homopolymers or also control the GC content. However, neither is done in this approach.

For the experiment, the message "JUNE6 INVASION:NORMANDY" was encoded, resulting in a sequence of 69 nucleotides. The sequences are flanked by two 20 nucleotide primers, resulting in an overall length of 109 nucleotides. Given that the same alphabet could be encoded with 6 bits per symbol, this yields an  information density of 1.27 b/nt. 

\subsection{Some Possible Codes for Encrypting Data in DNA~\texorpdfstring{\cite{Smith2003SomeDNA}}{}}\label{sse:smith}
Subsequent work~\cite{Smith2003SomeDNA} is the first to use non-trivial encoding and thus offers a considerable improvement in terms of information density. The authors propose three encoding schemes: \textit{Huffman codes}, \textit{comma codes} and \textit{alternating codes}. The development of the three schemes is driven by efficiency and robustness: the former is achieved by packing more information in the same number of nucleotides using compression and the latter by ensuring that insertions and deletions can be detected (to some degree). A further design goal is that  the output of the encoding should be clearly recognisable as artificial, so that it cannot be confused with DNA that occurs naturally. 

The approaches are theoretical, i.e., they have not been tested experimentally, and focus primarily on encoding the information. Although both the comma code and the alternating code possess basic inherent error detection, these approaches do not consider another important encoding factor---error correction, nor do they address incorporation of primers for amplification or the information retrieval method, i.e., sequential access vs random access.

\subsubsection{The Huffman Code}\label{sss:smith_huffman}
This approach follows the classic algorithm introduced by Huffman~\cite{huff1952amethod}. The input language is encoded such that the most frequent character is encoded with the least number of symbols, and similarly the least frequent input character is encoded with the most symbols (using variable-length encodings). For a given alphabet and language, this is the optimal symbol-by-symbol code. For the approach, the Huffman code is used to translate the letters of the Latin alphabet according to English language character frequencies, e.g.,

\begin{align}
    e &\rightarrow \texttt{T} \ &\text{(12.7\% frequency)} \nonumber\\
    n &\rightarrow \texttt{GC} \ &\text{(6.7\% frequency)} \nonumber\\
    f &\rightarrow \texttt{ACG} \ &\text{(2.2\% frequency)} \nonumber\\
    z &\rightarrow \texttt{CCCTG} \ &\text{(0.1\% frequency)} \nonumber
\end{align}

The most frequent letter in the English language is \textit{e}, and it is encoded as a single nucleotide, \texttt{T}. As the frequency decreases, the codeword grows in size. The average codeword length is 2.2 nucleotides. The size of the codewords ranges between one and five, and the size of the code is 26 (for the letters of the alphabet).

While this code is uniquely decodable and optimal in the sense of being the most economical for symbol-by-symbol coding, it also comes with three disadvantages. First, it is cumbersome and not optimal for encoding arbitrary data, and in the case of encoding text, the  frequency of letters depends heavily on the particular language. Second, due to the varying codeword length, it is difficult to observe a pattern in the encoded data, such that it could result in sequences similar to naturally occurring data. The avoidance of naturally occurring sequences could be an advantageous feature of an encoding scheme, allowing to easily differentiate artificial and natural DNA. Third, no error detection or correction codes are incorporated.
Additionally, the encoding scheme does not attempt to prevent homopolymers and suboptimal GC content. Given that only letters are considered, 5 bits suffice to encode each letter, this yields an overall information density of 2.27 b/nt. 

\subsubsection{The comma code}\label{sss:smith_comma}
This code uses one nucleotide, \textit{G}, as a comma to separate all other codewords with a length of five nucleotides but never uses it elsewhere (i.e., in the codewords themselves). The proposed code uses \texttt{G} as the comma occurring every six nucleotides: 

\begin{align}
    \texttt{G}*****\texttt{G}*****\texttt{G}*****\texttt{G}... \nonumber
\end{align}

The five-nucleotide codewords are limited to use the remaining three nucleotides: \texttt{A}, \texttt{C}, and \texttt{T}, with the added constraint that there must be exactly three \texttt{A} or \texttt{T} nucleotides and exactly two \texttt{C}s, in any order. This general format leads to 80 different possible codewords. Balancing \texttt{C}'s and \texttt{A} or \texttt{T} has the benefit that it balances the GC content (and thus leads to a more efficient amplification process). 

The main advantage of the approach is that it provides simple error-detection capabilities. Insertions and deletions can easily be detected: codewords which are too long or too short have clearly been subject to insertions and/or deletions. Changes, i.e., flipping of one nucleotide, can be detected in 83\% of the cases. Given a five-letter codeword with a one-letter comma, three possible point mutations can occur in each position, resulting in 18 single-point mutations. Of these single-point mutations, only 17\% result in valid codewords (flipping \texttt{A} to \texttt{T} or vice versa) and the remaining 83\% can be detected.
Using one \texttt{G} and two \texttt{C} nucleotides in one codeword means that the GC content is exactly 50\% and is thus well suited for sequencing. A final advantage is that the occurrence of a \texttt{G} nucleotide in every sixth position means that the sequence can easily be identified as synthetic. While this is not crucial for data storage, it is a design goal of the authors.
The disadvantages of the encoding are that no mechanism for error correction is provided. Also, DNA sequences produced with this encoding can contain homopolymers of length three.
The encoding uses 80 different codewords, each with a length of 6 nucleotides. To encode about the same number (128) of symbols with bits, 7 bits are required. The information density therefore is 1.17 b/nt.


\subsubsection{The alternating code}\label{sss:smith_alternating}
This code is made of 64 codewords of length 6, of the form \texttt{XYXYXY}, where $\texttt{X} \in \{\texttt{A}, \texttt{G}\}$ and $\texttt{Y}  \in \{\texttt{C}, \texttt{T}\}$ (the \texttt{X} and \texttt{Y} are alternating). The alternating structure is arbitrary and it is argued that other formats, e.g., \texttt{YXYXYX} or \texttt{XXXYYY} have the same properties. 

The advantages of this method are similar to the comma code: it has a simple structure and a ratio of 1:1 of the nucleotides, suggesting that it is an artificial code. It also shares the error-detection features of the comma code, although to a lesser extent (67\% of the mutations result in words that are not part of the code and hence can be identified as errors). Furthermore, given the suggested structure of \texttt{XYXYXY}, homopolymers of length three are avoided.
Disadvantages include a potentially suboptimal GC content, although this could be avoided by using a structure such as \texttt{XYXYXY} with $\texttt{X} \in \{\texttt{C}, \texttt{G}\}$ and $\texttt{Y}  \in \{\texttt{A}, \texttt{T}\}$, as well as the lack of error correction codes. 

The encoding uses 64 different codewords. To encode the same number of symbols in a binary representation, 6 bits are needed and thus the information density is exactly 1 b/nt.

\subsection{Alignment-Based Approach for Durable Data Storage into Living Organisms~\texorpdfstring{\cite{Yachie2007Alignment-basedOrganisms}}{}}\label{sse:yachie}
A subsequent project~\cite{Yachie2007Alignment-basedOrganisms} avoids explicit error correction codes. Instead, encoded messages are inserted into the genome of the host organisms repeatedly.

The message is first translated to the Keyboard Scan Code Set2~\cite{KeyboardCodes}. This hexadecimal code is then converted to binary, and a binary encoding to dinucleotides (pairs of nucleotides) is used to convert the bit sequence into a DNA sequence. The mappings are illustrated in Figure~\ref{fig:yachie}. The message encoded in the experiments is "E=mc\textasciicircum2 1905!".

\begin{figure}[h]
\begin{center}
\noindent\begin{minipage}[c]{.33\linewidth}
\begin{align*}
    \texttt{\%12\%24\%12\%4E\%3A}\\
    \texttt{\%21\%55\%1E\%29\%16}\\
    \texttt{\%46\%45\%2E\%12\%16}\\
\end{align*}
\end{minipage}%
\noindent\begin{minipage}[c]{.33\linewidth}
\begin{align*}
    \texttt{0001 0010 0010 0100 0001}\\
    \texttt{0010 0100 1110 0011 1010}\\
    \texttt{0010 0001 0101 0101 0001}\\
    \texttt{1110 0010 1001 0001 0110}\\
    \texttt{0100 0110 0100 0101 0010}\\
    \texttt{1110 0001 0010 0001 0110}\\
\end{align*}
\end{minipage}%
\begin{minipage}[c]{.33\linewidth}
\begin{align*}
    \texttt{AA = 0000   AG = 1000}\\
    \texttt{CA = 0001   CG = 1001}\\
    \texttt{GA = 0010   GG = 1010}\\
    \texttt{TA = 0011   TG = 1011}\\
    \texttt{AC = 0100   AT = 1100}\\
    \texttt{CC = 0101   CT = 1101}\\
    \texttt{GC = 0110   GT = 1110}\\
    \texttt{TC = 0111   TT = 1111}\\   
\end{align*}
\end{minipage}
\end{center}
\caption{Translation to Keyboard Scan code (left), translation to binary (middle) and the mapping from binary to dinucleotides (right).}
\label{fig:yachie}
\end{figure}

\begin{figure}[h]
    \centering
    \includegraphics[width=0.8\linewidth]{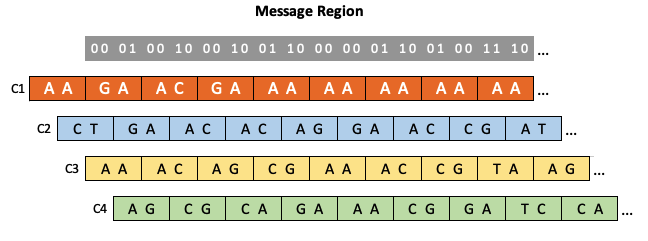}
    \caption{Illustration of the encoding of the same bit sequence, with different start and end bit sequences, four times.}
    \label{fig:yachie_strands}
\end{figure}

\noindent Redundancy is introduced by encoding the message four times. More precisely, the bit sequence is copied four times, with each copy flanked by different start and end bit sequences of different length. With the help of these bit sequences, the DNA sequences can be identified within the genome of the host organism. Owing to the different lengths of the start and end bit sequences, the resulting DNA sequences are distinct, and should a homopolymer occur, this can be filtered out and the other start-end bit sequences used instead. More specifically, for the first encoding (C1), the bit sequence is flanked with \texttt{0000} and \texttt{1111} and hence the DNA sequence is flanked with \texttt{AA} and \texttt{TT}. For the second encoding, the bit sequence is flanked with \texttt{111} and \texttt{0} and the DNA sequence thus starts with \texttt{GT} and ends with \texttt{AT}. The bit sequences for the third and fourth copy are flanked with (\texttt{00}, \texttt{00}) and (\texttt{1}, \texttt{111}), respectively. As the bit shift is less than four bits for the second, third and fourth copy, the messages are encoded in completely different DNA sequences. Figure~\ref{fig:yachie_strands} illustrates the different encodings.

These four encoded messages are then inserted in different places of the genome of the host organism (in this experiment, a strain of \textit{Bacillus subtilis}). To decode the information, the entire genome is sequenced and then searched for start and end sequences. For this procedure to work, the messages should be above a minimum chosen length. Otherwise, naturally occurring duplicate sequences might be mistaken for the encoded message. For this experiment and particular organism (\textit{B.subtilis}) it is argued that a minimum length of 20 provides enough specificity to avoid the messages to be mistaken for host DNA.

As with other techniques that store information in an organism, evolution and mutation pose problems, since the message might be altered. This problem is mitigated by using redundant copies placed in different regions of the genome, although this means the whole genome has to be sequenced, rendering the approach potentially inefficient. The authors predict that a large-scale system is possible, although as we explained, the size of the sequences in \textit{in-vivo} storage is very limited.

The approach does not control homopolymers or the GC content. Error detection and correction are also not taken into account, although redundancy helps to remedy errors. In the experiment carried out, however, the data was not completely recovered as only 99\% accuracy was achieved.

The encoding uses two nucleotides to encode four bits and thus suggests an information density of 2 b/nt. However, when taking into account that every message is copied four times, and that the start and end sequences are 5 bits on average across the copies, this leads to an overall information density of 0.49 b/nt.

\subsection{Length-Based Encoding of Binary Data in DNA~\texorpdfstring{\cite{portney}}{}}\label{sse:portney}
Subsequent work~\cite{portney} encodes information entirely differently with the goal of avoiding sequencing to read the information back. They accomplish this by taking the binary information and encoding a $0$ with a subsequence that has a different weight than the subsequence for $1$. This allows the content of the encoded information to be decoded based on the weight of the subsequences, making it possible to, for example, use gel electrophoresis instead of sequencing.  

In more practical terms, the subsequences representing binary $0$'s and $1$'s are separated by restriction sites which allow to cut or separate the encoded information precisely. The entirety of encoded information is inserted into the genome of a bacteria, also flanked by restriction sites (defined subsequences of biological relevance) such that it can be cut out precisely.

To facilitate decoding of the DNA oligonucleotides by electrophoresis, which identifies molecules by their masses, this approach uses 4 nucleotides to encode a $1$ and 8 nucleotides for a $0$. Information density thus depends on the exact message encoded. In the experiment the message 'MEMO' was encoded with a 3-bit alphabet---the payload of 12 bits (4 characters $\times$ 3 bits) was encoded in 60 nucleotides. The information density of the payload is, therefore, 0.2 b/nt. However, 50 additional nucleotides are required for restriction sites, which are essential for the encoding, resulting in a sequence of 110 nucleotides and the overall information density of 0.11 b/nt. This is very low and is mostly due to the need for restriction sites.

No error detection and correction codes are explicitly incorporated in the encoding; however, with decoding relying on the mass of subsequences, the encoding is resilient to substitutions of nucleotides (as this barely affects the mass of a subsequence) and may also be resilient to limited insertions or deletions (as there is a twofold difference in mass between the subsequences for $0$ and $1$). Although no particular reference is made to biological constraints, given the encoding uses a plasmid transfected into a host bacteria, care is needed to ensure that the sequences introduced do not disrupt cellular activity of the host. As only two different codewords are needed---one for $1$ and one for $0$---the respective subsequence can be chosen to (a) balance GC content and (b) avoid homopolymers. Regarding the former, even though the distribution of $1$'s and $0$'s may be uneven, as long as the GC content within the two subsequences is balanced, the GC content of the encoded sequence will be balanced as well.

Having each subsequence representing a bit flanked by restriction sites limits the scalability of the approach as the number of bits that can be stored depends on the number of unique restriction sites available. At the same time, however, using restriction sites enables, albeit rather coincidental, random access to the data. 

\subsection{An Improved Huffman Coding Method for Archiving Text, Images, and Music Characters in DNA~\texorpdfstring{\cite{Ailenberg2009AnDNA}}{}}\label{sse:ailenberg}
This is one of the first works to move beyond a basic proof of concept storing simple text messages. It archives text, image and music. The encoding is fundamentally based on an improved Huffman code, building upon previous methods such as the Huffman method described in Smith et al.~\cite{Smith2003SomeDNA} (Section~\ref{sss:smith_huffman}). The stored files are not in a standard digital format such as \texttt{.mp3}, \texttt{.png}, etc. Instead, music is encoded as a series of notes, with additional rhythm information, while for images, it is only possible to encode a series of primitives (circle, ellipse, line, rectangle) along with their position (and orientation).


The approach uses two components: the index plasmid and the information plasmid. The index plasmid is responsible for holding metadata such as title, authors, size of library and primer information. Similarly to Bancroft et al.~\cite{Bancroft2001Long-TermDNA} (Section~\ref{sse:bancroft}), the index information is flanked by unique primers, and thus is easy to find, amplify and decode. However, in contrast with~\cite{Bancroft2001Long-TermDNA}, the sequencing primers needed to find the actual data are not themselves stored in the index plasmid but in the information plasmid of the library---this is termed the iDNA in~\cite{Bancroft2001Long-TermDNA}, and we will reuse the term henceforth when the same method is used. Thus the index is responsible only for information and structure describing the actual data.

The approach encodes text, music and images. To distinguish between the type of content encoded, it is prepended by \textit{tx*}, \textit{mu*} and \textit{im*}, respectively.

One of the main advances proposed by this approach is a more general encoding based on the Huffman code. Previous methods have encoded a limited set of symbols---such as, for example, only letters without no digits or punctuation, or a very short message with symbols (albeit a small subset of keyboard scan codes) as in the encoding of "E=mc\textasciicircum2 1905!" by \cite{Yachie2007Alignment-basedOrganisms} (Section~\ref{sse:yachie}). However, the important factor is the potential number of available symbols an encoding scheme offers, rather than the number that is actually used in a demonstration encoding. In this approach, for text encoding, a complete keyboard input set (including shift, space and so on) is encoded by splitting the symbols into three groups of under 25 characters each. Each group has a header, denoted by one or few nucleotides. More specifically, Group 1 $\rightarrow$ \texttt{G}, Group 2 $\rightarrow$ \texttt{TT} and Group 3 $\rightarrow$ \texttt{TA}. The characters in each group are encoded in decreasing order of their frequency, as is common for Huffman codes. For example, the letter \textit{e} is encoded as \texttt{GCT}, where \texttt{G} denotes the group header and \texttt{CT} is the encoding for \textit{e}, \textit{shift} is \textit{GTC} and \textit{g} is \textit{GAAT} and encoding the word \textit{Egg} thus becomes \textit{GTCGCTGAATGAAT}.

For music, a single column is used to provide encodings for note values, pitches and meter. For example, a \texttt{D} half-note is encoded as \texttt{CGTT}, where \texttt{G} denotes the \texttt{D} and the \texttt{TT} indicates a half-note. Encodings are provided for all other notes. 

For images, the graphic primitives along with their properties, e.g., location, size and orientation are encoded. An excerpt of the encoding for graphic primitives is:

\begin{center}
    $; \rightarrow \texttt{G}$ \quad
    $. \rightarrow \texttt{TT}$ \quad
    $0 \rightarrow \texttt{TA}$ \quad
    $1 \rightarrow \texttt{AT}$ \quad
    $2 \rightarrow \texttt{CT}$ \quad
    ... \quad
    $S (s;x1;y1;a) \rightarrow \texttt{AAG}$  \quad
    $L (x1;y1;x2;y2) \rightarrow \texttt{CAA}  \quad$
    ...
\end{center}

A square (S) is defined by the length of its sides (s), the location of its upper right vertex (x1,y1) and the angle of its base (a). A line (L) is defined by its two endpoints (x1,y1) and (x2,y2).

For example, a square in location (0,0) with a side of 1 and angle of 0 will be encoded as \texttt{AAGATTATATA}. Regardless of how many graphic primitives are used, the first one has to be prepended with \textit{im*} or \texttt{GCGTAACTACCA} to indicate the start of graphic primitives.

In theory, an iDNA library of size up to 10,000 nt could be achieved by introducing sequencing primers (of length 20-30 nt) every 500-nt fragment. This limitation of library size occurs due to a limit on the amount of non-organism DNA that the plasmid is able to tolerate while still allowing for the successful amplification. The paper provides an example where 409 nt of text, 113 nt of music, and 238 nt of image data, with an additional 16 nt for punctuation and 68 nt for three primers, for a total of 844 nt used. The information was retrieved with 100\% accuracy, although the authors suggest that sequencing with the three primers should be done in a particular order (so as to minimise the chance of retrieving some of the plasmid's DNA).

This method comes with the common advantages and disadvantages of storing information in organisms capable of independent replication (\textit{plasmids}). While this is the first work to store considerable amounts data of different kinds (text, music and image data), the implementation is only capable of storing simple, structured data (discrete music notes, graphical primitives) and not actual recordings or photographs. 

The encoding itself uses a Huffman code for characters and music or graphic primitives. For encoding text, given the distribution/frequency of characters/symbols in the English language, the encoding requires on average 3.5 nucleotides for the each of the 71 symbols used. The least number of binary bits required to encode the nearest number of characters (128) is 7 bits, resulting in an information density of 2 b/nt. Assessing the information density of music and graphic, however, is not possible as the distribution of the different primitives and numbers (coordinates, length etc.) is not known.

Generally, the encoding does not take into account biological constraints like GC content or homopolymers, nor does it provide a mechanism for error correction and detection. On the positive side, the encoding is very economical and uses very few nucleotides per bit, at least for text. 


\subsection{Blighted by Kenning~\texorpdfstring{\cite{kenning}}{}}\label{sse:kenning}
This project uses a very simple mapping where one letter of the alphabet is mapped to three nucleotides, e.g., \texttt{O} $\rightarrow$ \texttt{ATA}. The project has primarily artistic purposes. The encoding is used to map article 1 of the Universal Declaration of Human rights to synthetic DNA. The DNA is then inserted into a bacteria sprayed on apples (to represent the forbidden fruit as well as the tree of knowledge).

With primarily artistic purposes, the project does not take into account error detection/correction or biological constraints. It requires full sequencing to retrieve the information and is stored in a bacterial genome which was then sprayed onto apples. Given that three nucleotides are used to encode a letter, an alphabet of 64 letters can be encoded, which results in 6 bits per letter and  an information density of exactly 2 b/nt. 

\section{Advanced Approaches}\label{sec:advanced}
In this section, we cover modern approaches to DNA data storage.
In contrast to much of the earlier work on DNA storage, the approaches in this section focus on encoding arbitrary bit streams instead of specific file types (e.g., text).
(An exception is the work~\cite{dimopoulou2020image} discussed in Section~\ref{sse:jpeg} that provides an advanced method for storing JPEG images.)
They often feature more advanced error correction and random access, and all of them store the synthesised DNA on a microplate.

\subsection{Next-Generation Digital Information Storage in DNA~\texorpdfstring{\cite{Church2012Next-generationDNA}}{}}\label{sse:church}
This work by Church et al. is an important milestone in DNA storage, as it is the first to store relatively large amounts of information (5.27 MB). This is possible due to the use of "next-generation" (at the time) sequencing and synthesis technologies. The goal is long-term, archival storage. The archived content is composed of a 53,426-word draft of a book, 11 JPG images and one JavaScript program. Instead of being inserted into a living organism, the synthesised DNA is simply stored on a microplate.

The approach has multiple advantages, starting with the encoding: a nucleotide encodes one bit (\texttt{A} or \texttt{C} for zero, \texttt{G} or \texttt{T} for one), instead of two. 
It is thus possible to encode the same message (as a sequence of bits, i.e., 0s and 1s) as different sequences of nucleotides. A bit sequence of \texttt{1111} can, for example, be encoded as \texttt{GGGT}, thus avoiding a homopolymer of length greater than three. The decision whether a zero is encoded as \texttt{A} or \texttt{C} (or a one as \texttt{G} or \texttt{T}) is made at random unless a homopolymer of length four needs to be avoided. The randomness automatically ensures a somewhat balanced \texttt{GC} content (although it could also be enforced).

The information is split into addressable data blocks (which are encoded individually) to avoid long sequences, which are generally difficult to synthesise and which may also form secondary structures. Each block has a 19-bit code, or, due to mapping one bit to one nucleotide, a 19-nucleotide DNA subsequence that identifies its position in the data. Each sequence is flanked by primers of length 22 for amplification when sequencing.

Synthesis and sequencing errors have a low probability of being coincident, e.g., errors are unlikely to occur in the same location in different copies of the sequence. To address error correction and detection, each sequence is consequently synthesised multiple times, enabling errors to be identified easily by comparing multiple noisy reads of the sequence, i.e., by computing a consensus via a majority vote.

\begin{wrapfigure}{L}{0.75\textwidth}
\centering
\includegraphics[width=0.7\textwidth]{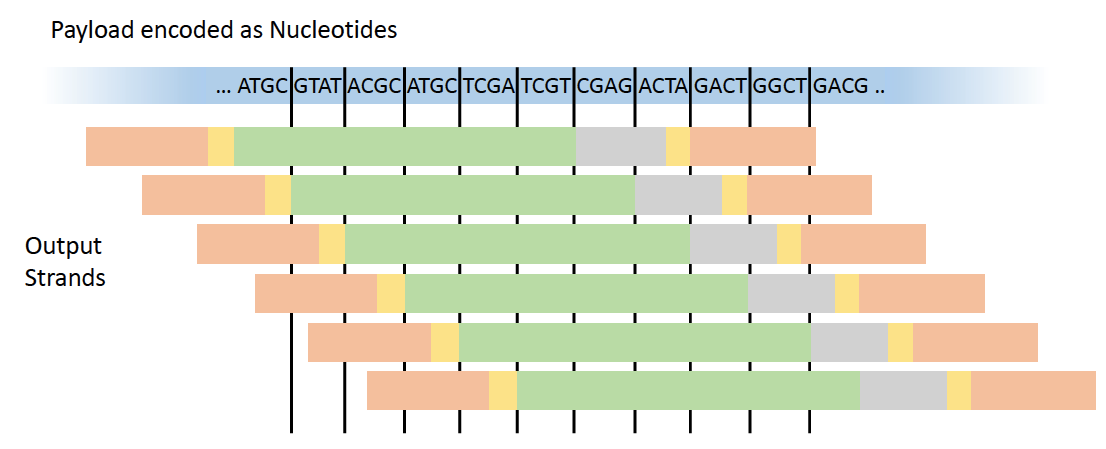}
\caption{The Goldman encoding uses fourfold redundancy of each segment.}
\label{fig:goldman}
\end{wrapfigure}

This encoding uses a single nucleotide to encode each bit and achieves a physical storage density of 5.5 petabits/mm$^3$, reinforcing the idea that DNA is suitable as a storage medium for archival data. The one shortcoming is the lack of a more sophisticated error detection and correction mechanism. With the simple encoding, a number of small errors (10, mostly caused by homopolymers) occurred, indicating that a more robust technique is needed. The reported long access time (hours to days) is on par with other approaches (due to the slow speed of sequencing) and is acceptable considering this is an approach for archival data which is accessed only very infrequently.

The theoretical payload-only information density of this approach is 1 b/nt. While this is not particularly high---2 b/nt is the theoretical maximum achievable without the use of compression---this encoding approach both avoids homopolymers and also allows to balance the GC content. Taking into account the addressing information (19 nucleotides) and the sequencing primers (two primers of length 22 nucleotides each), only 96 nucleotides can be used to store data (given the sequence length of 159 nucleotides that was used in the experiments). This results in an information density of 0.6 b/nt. 
However, as no error correction codes are used, each sequence needs to be replicated multiple times and sequenced with a high coverage (a coverage of approximately 3000 is used; however, no attempt has been made to perform a sub-sampling experiment to find a minimum coverage that would be sufficient).

\subsection{Towards Practical, High-Capacity, Low-Maintenance Information Storage in Synthesized DNA~\texorpdfstring{\cite{Goldman2013TowardsDNA}}{}}\label{sse:goldman}
Subsequent work by Goldman et al.~\cite{Goldman2013TowardsDNA} also encoded a significant amount of binary data (comprising text, PDF documents, MP3 audio files and JPEG images), totalling 757,051 bytes.
Each of the bytes of the files is mapped to base-3 numbers. A Huffman encoding is used to map each byte to either five or six base-3 numbers, which in turn are translated to nucleotides. More specifically, each base-3 number is translated to a nucleotide, picked such that it is different from the previous one to avoid homopolymers.

The resulting long sequence is partitioned into segments of 100 nucleotides, each overlapping the previous by 75 nucleotides, thereby introducing fourfold redundancy since every 25-nucleotide subsequence will be in four sequences as illustrated in Figure~\ref{fig:goldman}. To map each sequence to a file, two base-3 numbers are used, allowing for 9 different files to be stored. To order the sequences within a file, 12 base-3 numbers are used, allowing for 531,441 locations per file. One base-3 number is used as parity check and appended at the end (encoded in the same way as the payload so that no homopolymer is created).

The fourfold redundancy provides effective error correction: as each nucleotide is encoded in four DNA segments, most errors in synthesis or sequencing can be corrected by a majority vote. 
The approach is specifically designed to avoid homopolymers, but no specific mechanism is used to balance the GC content. Error detection is first implemented using the parity nucleotide. The fourfold redundancy then helps to correct errors by a majority vote. The approach is based on Huffman encoding, so the information density depends on the data encoded as some bytes are mapped to 5 and some bytes to 6 base-3 numbers. For the data used in the experiments, taking into account both compression and fourfold redundancy, an overall information density of 0.22 b/nt is achieved.

\subsection{Robust Chemical Preservation of Digital Information on DNA in Silica with Error-Correcting Codes~\texorpdfstring{\cite{Grass2015RobustCodes}}{}}\label{sse:grass}
Grass et al. propose a novel approach that incorporates error detection and correction. With it, information can be reliably extracted from DNA that is treated to simulate a 2000-year storage period in appropriate conditions. The amount of information stored, 83 KB organised into 4991 DNA segments of length 158 nt each (with 117 nt used to encode information), is much less than in~\cite{Church2012Next-generationDNA} (Section~\ref{sse:church}), but the focus is on the novel encoding, embedding the DNA in silica, and the simulating of ageing of the DNA for long-term storage.

The encoding mechanism is based on Reed-Solomon (RS) codes, a group of well-studied error-correcting codes which have the property of being \textit{maximum distance separable} (MDS). This property is important because an MDS code with parameters \textit{(n, k)} has the greatest error detection and correction capabilities compared to other codes with the same parameters. The set of alphabet symbols of an RS code is a finite field and the number of elements must therefore be a power of a prime number.

\begin{wrapfigure}{L}{0.45\textwidth}
\centering
\includegraphics[width=0.40\textwidth]{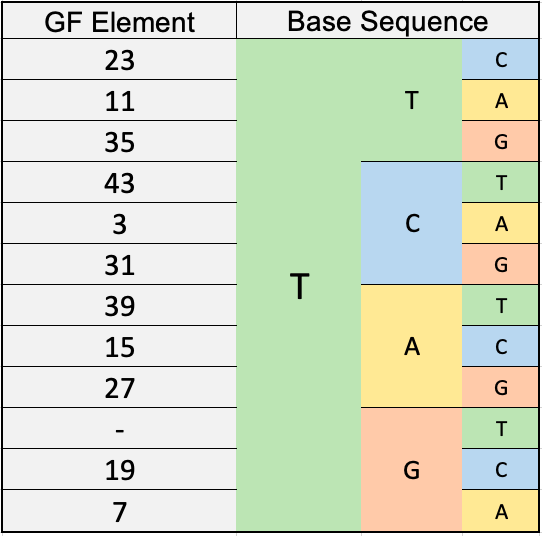}
\caption{The codon wheel translates base 47 numbers to DNA sequences of length three.}
\label{fig:codon}
\end{wrapfigure}

\begin{figure}[h]
    \includegraphics[width=0.96\linewidth]{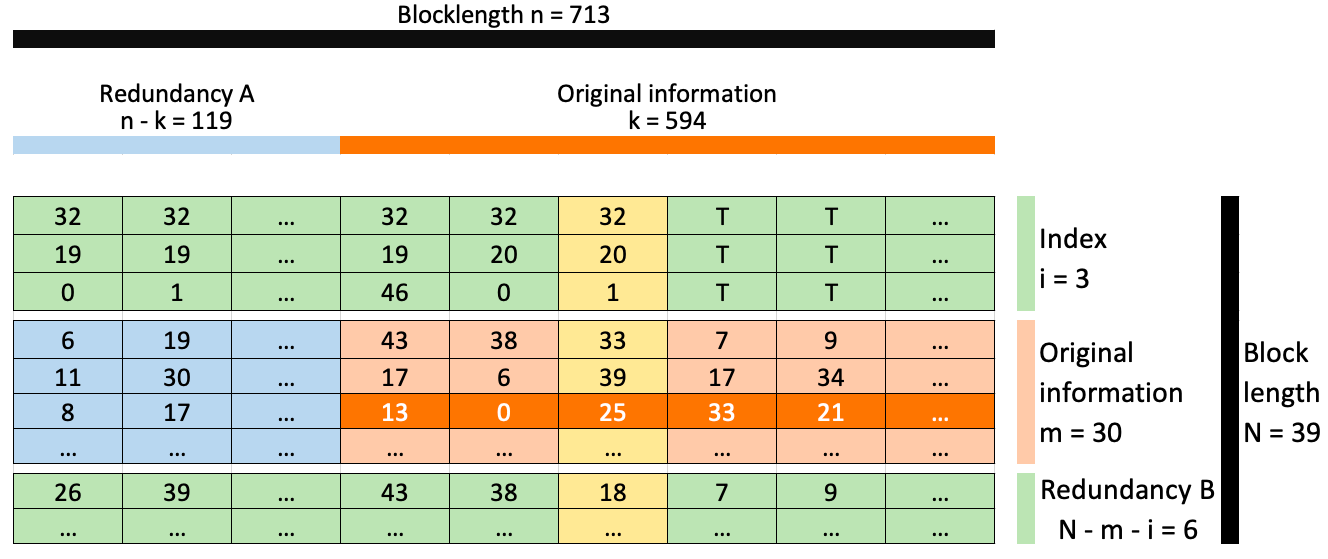}
    \caption{Encoding approach in~\cite{Grass2015RobustCodes}: indexing as well as inner and outer error correction coding applied to data blocks.}
    \label{fig:grass_encoding}
\end{figure}

Typical input data are files, which can be thought of as sequences of bytes or of a number in base $2^8 = 256$. The first step of the algorithm is to convert two bytes of the input to three numbers in the finite field $GF(47)$ (called a Galois Field, hence GF). The number 47 is chosen as it is the closest prime number to 48, the number of 3-nucleotide sequences that can be constructed to satisfy biological constraints such as avoiding homopolymers. Once the conversion to $GF(47)$ is done, the resulting numbers are arranged into a $30 \times 594$ block. The information words are shown in Figure~\ref{fig:grass_encoding} in the red box. The dark red box shows one row of such information words. Then RS codes are used on the \textbf{rows} to add redundancy information of size 119 (shown in the columns on the left), for a total block length of $594 + 119 = 713$. This is called the \textit{outer} RS code. In the next step, an index of size 3 is added to each \textbf{column} (shown at the top), for a vector of size 33, followed by the use of a second (\textit{inner}) RS code on the resulting length-33 vector that adds an extra 6 numbers (shown at the bottom), thus producing a column of the total size 39 and resulting in a block of size $39 \times 713$. The index serves to identify each column (one column will be stored in one DNA sequence) and, given the length of 3 in base 47, allows to address 103,823 columns. The inner RS code is added to correct individual nucleotide errors within sequences, whereas the outer RS code operates across sequences and may potentially recover completely lost sequences.
The columns (one is shown as the yellow rectangle in Figure~\ref{fig:grass_encoding}) are uniquely mapped to DNA sequences of length 117, by mapping each element/number to three nucleotides, as described above. Primers, referred to as sequencing adapters, are added to flank the resulting sequence.

For the decoding process, the DNA is sequenced using the Illumina MiSeq platform. On the sequences read, the inner code is decoded followed by the decoding of the outer code. In the experiments the inner code fixes on average 0.7 nucleotide errors per sequence while the outer code fixes on average the loss of 0.3\% of total sequences and corrects about 0.4\%. The information was completely recovered, without errors. Furthermore, an experiment where the DNA suffers decay (achieved by thermal treatment) equivalent to one and four half-lives (about 500 and 2000 years, respectively) shows that the encoding scheme, aided by half-life-extending storage in silica, can retrieve the information without errors even in these conditions.

This work is the first to incorporate error detection and correction codes without resorting to replication. In this encoding, the RS codes are chosen such that they can fix 3 arbitrary errors per sequence, correct 8\% of sequences if they are incorrect and recover 16\% of all sequences if they are missing. Homopolymers and GC content are also incorporated through the use of carefully picked subsequences (in the so-called codon wheel shown in Figure~\ref{fig:codon}).
The work encodes 83 kilobytes into 4,991 sequences of length 158, resulting in an overall information density of 0.86 b/nt.

\subsection{A DNA-Based Archival Storage System~\texorpdfstring{\cite{Bornholt2016ASystem}}{}}\label{sse:bornholt}
The major novelty of the work~\cite{Bornholt2016ASystem} by Bornholt et al. is the concept of random access, a feature missing from previous approaches. While the goal is still archival storage, with read-write times similar to other implementations (hours to days), the addition of random access enables a new level of efficiency, as the sequencing of all DNA  is no longer needed.
Previous work stored the DNA in a single pool, which comes with the disadvantage that the whole pool needs to be sequenced to retrieve even just a subset of the data. Using a separate pool for each object represents too big of a sacrifice in storage density. To enable retrieval of different subsets of data, i.e., files, from a single pool, the novel solution proposed by the authors is to attach different primers to each sequence, thus enabling their individual retrieval.
When writing, the primers are added to the DNA sequence. At read time, the primers are used for selective PCR amplification of only the desired data. As DNA molecules do not have a particular spatial organisation, each encoded sequence must contain an address that identifies its position in the original data.

\begin{figure}[h]
\centering
\includegraphics[width=0.88\textwidth]{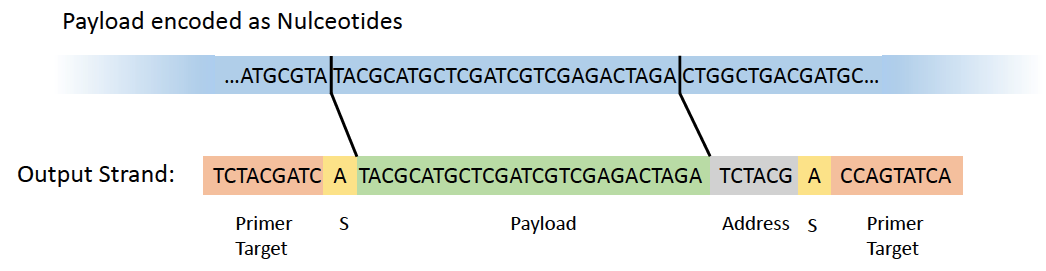}
\caption{Sequence structure containing primer, address and payload.}
\label{fig:bornholt_strand}
\end{figure}

As before, to encode a large amount of information, the data is split into blocks, which have the following structure: \textit{primers} at both ends, a \textit{payload} of information flanked by two \textit{sense} nucleotides that aid the decoding, and an \textit{address} (see Figure~\ref{fig:bornholt_strand}).
The \textit{payload} represents the actual data to be stored. As mentioned, this is a fragment of the entire information that needs to be encoded, and its length can be adjusted depending on the chosen sequence length. The two \textit{sense} nucleotides specify whether the sequence has been reverse complemented, which is helpful during decoding. The \textit{primers} are used, as usual, in the PCR amplification process to select the desired blocks and sequences. The \textit{address} indexes the sequence/block within the file stored.

As there is a unique mapping from files to primer sequences, all sequences of a particular file will share a common primer (because an object is associated with a single key). Thus the read operation is simply a PCR amplification using that particular file's primer. The sequences retrieved for a specific object can then be rearranged to recover the original information based on their \textit{address}. 

\begin{wrapfigure}{L}{0.53\textwidth}
\centering
\includegraphics[width=0.48\textwidth]{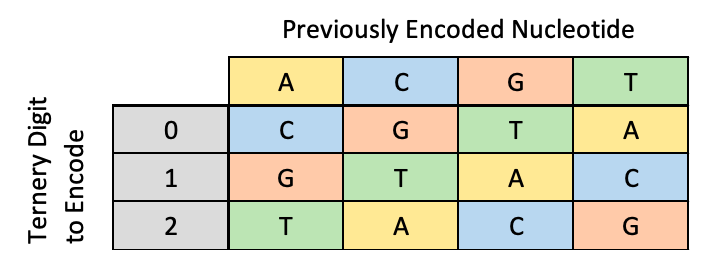}
\caption{Mapping of ternary codes to nucleotides.}
\label{fig:bornholt_mapping}
\end{wrapfigure}

The encoding from binary information to nucleotides is similar to previous approaches. Bytes are transformed using a ternary Huffman code to 5 or 6 ternary digits. The ternary strings are then converted to nucleotides using a rotating code that avoids homopolymers (the encoding for a 0, 1 or 2 depends on the previous nucleotide, such that sequences of repeating nucleotides are avoided). The principle of the mapping is illustrated in Figure~\ref{fig:bornholt_mapping}. For example, the binary string \texttt{01100001} is mapped to the base 3 string \texttt{01112}, which can be encoded to DNA as \texttt{CTCTG}. A Huffman code is used as an intermediary step, to map more common characters to 5 ternary digits, while less frequent ones are mapped to 6 digits.
This creates a degree of compression for textual data, although the authors report that the effect of such compression on the overall information density is insignificant.

\begin{wrapfigure}{R}{0.65\textwidth}
\centering
\includegraphics[width=0.6\textwidth]{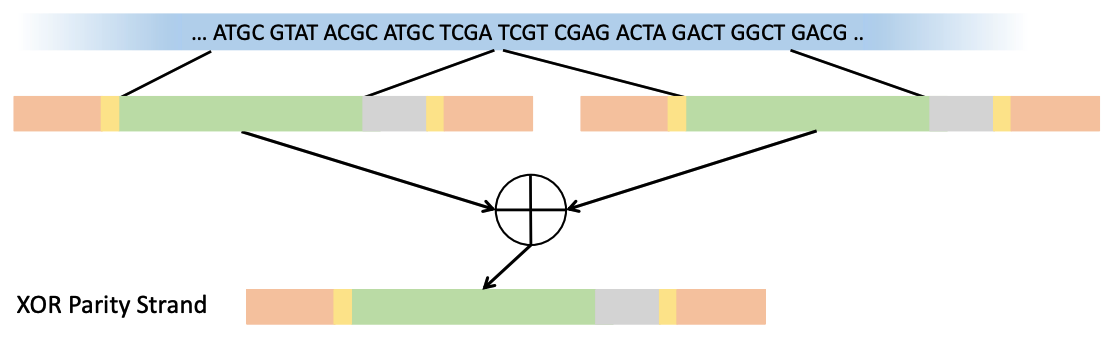}
\caption{The XOR of payloads of different sequence is used in a parity sequence to allow for error correction in~\cite{Bornholt2016ASystem}.}
\label{fig:xor}
\end{wrapfigure}

A novel way to provide error detection and correction through redundancy is introduced as well. The inspiration comes from the encoding used by Goldman et al.~\cite{Goldman2013TowardsDNA} (Section~\ref{sse:goldman}). Essentially, the Goldman encoding splits the DNA nucleotides into four overlapping fragments, thus providing fourfold redundancy (see Figure~\ref{fig:goldman}). This encoding is taken as a baseline to evaluate the results of the current scheme. Bornholt et al. propose an \textit{XOR Encoding}, in a similar fashion to RAID 5. With this approach, the exclusive-or operation is performed on the payloads $A$ and $B$ of two sequences, which produces a new payload $A\oplus B$. The address of the newly created block can be used to tell if it is an original sequence or the result of the exclusive-or operation. Any two of the three sequences $A$, $B$ and $A\oplus B$ are sufficient to recover the third. The major advantage of this method is that is offers similar reliability to the Goldman encoding, while being more efficient in terms of information density. The Goldman encoding repeats each nucleotide (up to) four times, while for the Bornholt encoding each nucleotide is repeated 1.5 times on average (see Figure~\ref{fig:xor}).

Overall, this design introduces a remarkable feature, \textit{random access}, as a proof of concept. It incorporates error detection and correction and avoids homopolymers. The information density of the payload only using this encoding is 1.05 b/nt. Accounting for the primers and addresses yields an overall information density of 0.59 b/nt.

\subsection{Portable and Error-Free DNA-Based Data Storage~\texorpdfstring{\cite{HosseinTabatabaeiYazdi2017PortableStorage}}{}}\label{sse:yazdi}
Another work that implements random access via PCR amplification has been developed concurrently by Yazdi et al.~\cite{HosseinTabatabaeiYazdi2017PortableStorage}. The approach (or rather experiments) use rather long DNA sequences of 1,000 nucleotides, of which 16 are used for address/index and the remainder is used for encoded payload.

The encoding of the payload and the design of the addresses follow three goals. First, none of the (distinct) addresses or indexes should be a part/subsequence of any of the encoded payload (weak mutual correlation). Second, the addresses should be as distinguishable as possible. The third goal is to balance GC content and to avoid homopolymers.

The developed method generates two sets of codewords, $S_1$ and $S_2$. $S_1$ is to be used as addresses, and the Hamming distance is used to make sure they are sufficiently different from each other. $S_2$ contains codewords to be used to encode the payload. Properties such as weak mutual correlation, a balanced GC content and the absence of homopolymers are provable with this encoding.

In their experiments, the authors encode the image of a black and white movie poster (Citizen Kane) as well as a colour image of a smiley face. The information is first compressed and then encoded in 17 sequences---each with a length of 1,000 nucleotides---by picking an address from $S_1$ and mapping the binary content of the images to codewords from $S_2$.

Errors are detected and corrected during post-processing, i.e., after sequencing. Using PCR before sequencing means that each sequence will be replicated multiple times. High-quality reads are identified as the ones where the address contains no errors. Then the high-quality reads are aligned and a consensus is computed. The alignment uses its knowledge of possible codewords as a hint. In the experiments, most errors at this stage are errors of deletions in homopolymers of length at least two. The remaining errors are further corrected by taking into account GC balance, which helps to determine which homopolymers are likely. A constraint is applied to subsequences of 8 nucleotides that ensures that the sum of the nucleotides, and any homopolymers, must be 8, which aids in the detection of deletions. To balance the GC content, the nucleotide subsequence is converted to binary and is selected if the word is balanced, i.e., it contains as many zeros as ones.

In summary, this encoding avoids homopolymers and also balances GC content. It does not use any explicit error correction codes and relies instead on sets of codewords that satisfy a number of constraints. Errors are detected and corrected in post-processing. In terms of information density, the compressed images are of size 29,064 bits and are encoded with 16,880 nucleotides (16 sequences of length 1,000 and one sequence of length 880), thus resulting in an overall information density of 1.72 b/nt. As is done in the work by Church et al. \cite{Church2012Next-generationDNA}, this method uses replication by PCR to correct errors when sequencing. More specifically, each sequence is amplified multiple times through PCR and subsequently sequenced. Errors in sequencing are detected by comparing the sequencing results of multiple replicas of the same sequence. The corrected sequence  is computed through a majority vote.

\subsection{OligoArchive: Using DNA in the DBMS storage hierarchy~\texorpdfstring{\cite{oligoarchive}}{}}\label{sse:oligoarchive}
Subsequent work~\cite{oligoarchive} focused on encoding structured database information and implementing database operations. Two different encodings are discussed, one for storing structured database information in the form of relational tables, and one for implementing database operations.

\begin{wrapfigure}{L}{0.65\textwidth}
\centering
\includegraphics[width=0.6\textwidth]{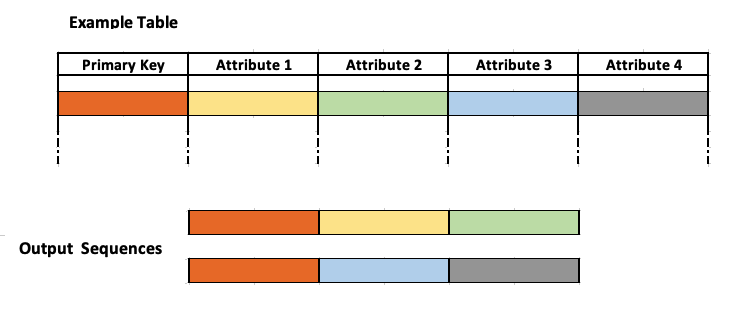}
\caption{Shredding a table into multiple sequences each containing the primary key.}
\label{fig:encoding1}
\end{wrapfigure}

The first encoding exploits the inherent structure in databases, i.e., that each attribute of a record in a table can be linked to the record  using the primary key. Doing so means that attributes of the same record can be distributed across different DNA sequences without the need for addressing. Instead the primary key is used for this purpose, reducing the space needed for the address. Figure~\ref{fig:encoding1} illustrates how one record of a table is stored in multiple sequences.

More specifically, dictionary encoding is used to compress the information. The dictionary is encoded in DNA as well. Subsequently, as many attributes as possible are stored in a DNA sequence along with the primary key (to link together attributes of the same record). A parity nucleotide is added to each DNA sequence for error detection; furthermore, each strand is duplicated, mirrored and reverse-complemented to provide additional error protection. After sequencing, the parity nucleotide and length of the DNA sequence are used to discard invalid sequences. The remaining sequences are aligned to compute a consensus by a majority vote. In the experiments, based on a subset of the database benchmark TPC-H\cite{TPC-H}, multiple tables are encoded, synthesized, sequenced and fully recovered.

The encoding used is based on previous work by Goldman et al.~\cite{Goldman2013TowardsDNA} discussed in Section~\ref{sse:goldman} and thus avoids homopolymers but does not guarantee a balanced GC content. Error detection is possible through the parity nucleotide but correction is left to the decoding process (through computing the consensus and pairing the strands with their mirrored and reverse-complemented copies).
The overall information density achieved on a 12-KB test database is 1.79 bits per nucleotide (the figure includes dictionary compression).

\begin{wrapfigure}{L}{0.65\textwidth}
\centering
\includegraphics[width=0.6\textwidth]{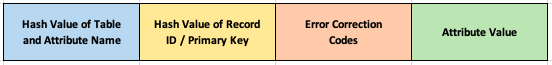}
\caption{Structure of the encoding.}
\label{fig:encoding2}
\end{wrapfigure}

\begin{wrapfigure}{L}{0.65\textwidth}
\centering
\includegraphics[width=0.6\textwidth]{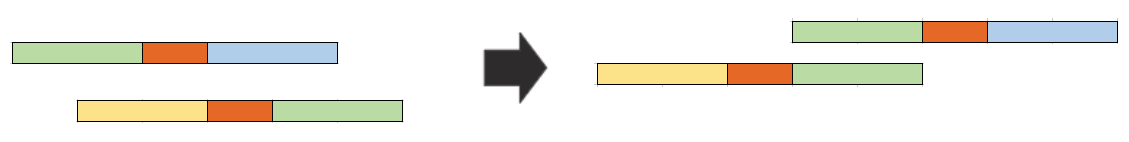}
\caption{Overlap extension PCR used to find matching tuples/sequences.}
\label{fig:oepcr}
\end{wrapfigure}

The second encoding developed in the same work enables performing both selective retrieval of records based on their value and database joins, i.e., finding pairs of records that agree on a particular attribute.
To enable these operations, each attribute is stored in one DNA sequence. Fixed-sized hash values are computed for variable-length fields. In particular, table name, attribute name and the primary key are hashed to obtain fixed-length strings and arranged in a DNA sequence together with the attribute value and parity-check symbols as shown in Figure~\ref{fig:encoding2}.
PCR amplification is used to retrieve all DNA sequences encoding a particular value for a specific attribute. Similarly, overlap extension PCR is used to implement a join by annealing matching sequence/attributes together as shown in Figure~\ref{fig:oepcr}.

In the case of both value-based retrieval and joining, the encoding of the value must also serve as a primer and must therefore be designed with extra care to avoid retrieving or joining records with values that are similar but not identical to the prompt. To that end, similar but non-identical values must become substantially different after encoding. This is achieved by encoding the value together with its SHA3 checksum: the values of an attribute $a$ and its checksum $c$ are split into subsets ($a_1$, $a_2$, $a_3$, \ldots and $c_1$, $c_2$, $c_3$, \ldots) and interleaved (resulting in a sequence $a_1$, $c_1$, $a_2$, $c_2$, $a_3$, $c_3$,  \ldots). Thanks to the avalanche effect of SHA3 (and other cryptographic functions)---small differences in input value lead to considerable differences in the checksum---the resulting sequences will be substantially different even if the attribute values are similar.

The second DB operations-oriented part of the project uses a straightforward mapping of two bits to one nucleotide.
This neither avoids homopolymers nor balances the GC content.
Error correction and detection is incorporated using a Reed-Solomon code.
With the hash and error-correction overhead taken into account, the resulting information density is 1.43 bits per nucleotide; since the primers are payload-dependent and no addressing is needed, the overall information density and the payload-only information density are equal.

\subsection{Image storage in DNA using Vector Quantization~\texorpdfstring{\cite{dimopoulou2020image}}{}}\label{sse:jpeg}
This work focuses on efficiently encoding JPEG images. The authors develop an encoder that is tailored to this specific input type and propose an end-to-end encoding workflow that optimally compresses an image before it is stored into DNA. This allows to control the encoding rate and potentially reduce the cost of synthesis. 

\begin{figure}[ht]
    \centering
    \includegraphics[width=0.95\linewidth]{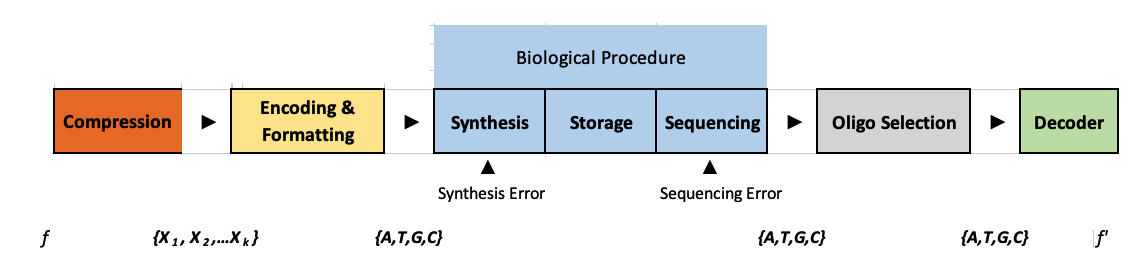}
    \caption{Proposed JPEG encoding scheme.}
    \label{fig:jpeg_encoding}
\end{figure}

The workflow consists of four stages, depicted in Figure~\ref{fig:jpeg_encoding}. The first stage performs image compression using discrete wavelet transform with each wavelet sub-band quantised independently using vector quantisation. The second stage performs the DNA encoding of the binary data into a quaternary code. The code is constructed using two dictionaries, C1 and C2 so as to regulate GC content and prevent homopolymer runs.

\begin{align}
C1 = {AT; AC; AG; TA; TC; TG; CA; CT; GA; GT}\\
C2 = {A; T; C; G}
\end{align}

C1 comprises two-nucleotide pairs and C2 contains single nucleotides, allowing even-length codewords to be generated using C1 and odd-length codewords longer than two nucleotides to be generated using a combination of C1 and C2. The third stage, known as formatting, involves cutting the generated DNA code into oligonucleotides of length 300 nt or less to conform to the limitations of current DNA synthesis technologies. After the DNA is stored, the fourth and final stage combines oligonucleotide selection and application of the decoder to retrieve the data.

A major advantage of the proposed approach is the use of an encoding algorithm that avoids homopolymers and balances the GC content, thereby making it robust to errors in the synthesis and sequencing processes. No additional measures are taken and so no error detection and correction codes are included.

A variable compression factor applied to the JPEG image before the encoding results in a range of information densities. We have assessed the payload-only information density achieved for the data that was actually encoded and found it to be 1.6 b/nt.
Since no wet lab experiment has been carried out, we do not specify the overall information density or coverage for this approach in Table~\ref{table:comparison}.

\subsection{Random Access in Large-Scale DNA Data Storage~\texorpdfstring{\cite{Organick2018RandomStorage}}{}}\label{sse:organick}
In this work by Organick et al.~\cite{Organick2018RandomStorage}, 35 distinct files, totalling over 200 MB, are encoded into a pool of over 13 million 150-nt sequences. One aim of the work is error-free recovery of the encoded files using Reed-Solomon codes. Another is to perform PCR-based random access, possible in principle in early schemes by Bancroft et al.~\cite{Bancroft2001Long-TermDNA} (Section~\ref{sse:bancroft}) and Church et al.~\cite{Church2012Next-generationDNA} (Section~\ref{sse:church}) and studied explicitly in later works by Bornholt et al.~\cite{Bornholt2016ASystem} (Section~\ref{sse:bornholt}) and Yazdi et al.~\cite{HosseinTabatabaeiYazdi2017PortableStorage} (Section~\ref{sse:yazdi}).

Using a similar approach to Bancroft et al.~\cite{Bancroft2001Long-TermDNA} (Section~\ref{sse:bancroft})---which used information DNA (iDNA) and polyprimer key (PPK) sequences---they partition each file's data into chunks, each of which is assigned to a large number of oligonucleotides that share the same primer (representing the file ID), and ordered using a sequence-specific ID. A subset of primers are selected from a larger pool by and discarding sequences which are pairwise too similar (based on the BLAST tool~\cite{blast}) as this may lead to retrieval of the wrong data, i.e., sequences with a similar primer. To avoid multi-bit repeats and make the sequences less similar, the input data for the files are subjected to XOR with a pseudo-random number sequence. A ternary code is then used to map the bytes to the DNA sequence with no homopolymers and a Reed-Solomon code is applied to the payload for error correction.

To retrieve the data from the DNA, the decoder first clusters oligonucleotides based on their content. Then a variation of the bitwise majority alignment algorithm~\cite{bma} is used to identify insertions, deletions and substitutions and to recover the original sequence via a consensus for each nucleotide position of the aligned reads. After Reed-Solomon decoding, an XOR with the same pseudo-random number sequence that was used during encoding yields the original data.

The authors have demonstrated that 200.2 MB of data (35 files) can be correctly recovered without any errors (byte-for-byte equal to original) from DNA, with an overall information density of 0.81 b/nt.
Notably, this method has experimentally achieved the lowest minimum coverage required to successfully recover the data among all surveyed approaches.

\subsection{DNA Fountain Enables a Robust and Efficient Storage Architecture~\texorpdfstring{\cite{erlich2017dna}}{}}\label{sse:erlich}
Another approach to encoding information, proposed by~\cite{erlich2017dna}, uses fountain codes~\cite{mackay2005fountain} to encode the data. In this work, 2,146,816 bytes (2.14 MB) of compressed data are encoded, including a complete graphical operating system of 1.4 MB and a 52-second movie.

The encoder comprises three steps: The first partitions the data into 32-byte chunks (although this can be a user-defined length). The second uses a pseudo-random number generator to select 32-byte chunks to which a Luby transform is applied. This is to produce what is referred to as droplets in the context of fountain codes---i.e., linear combinations of data chunks. This step involves bitwise addition of randomly selected subsets of chunks modulo 2, attachment of a random seed (protected with a Reed-Solomon code), which allows the decoder to identify which chunks were used to produce the droplet, and conversion of the resulting binary droplet into a DNA oligonucleotide whereby the binary \{00,01,10,11\} is mapped to \{A,C,G,T\}, respectively. The third step screens the droplet oligonucleotides such that they are filtered to conform to the biological constraints, i.e., have a balanced GC content and no homopolymer runs. Their approach achieves a payload-only information density of 1.75 b/nt and an overall information density of 1.19 b/nt.

\subsection{High Capacity DNA Data Storage with Variable-Length Oligonucleotides Using Repeat Accumulate Code and Hybrid Mapping~\texorpdfstring{\cite{wang2019high}}{}}\label{sse:wang}
The fountain-code approach in~\cite{erlich2017dna} (Section~\ref{sse:erlich}) requires storing the random seed used to generate a droplet, which lowers the overall information density.
On the other hand, storing the seed allows the encoder to simply filter out the droplets that do not conform to biological constraints.
In~\cite{wang2019high}, the authors propose an alternative two-step approach that does not require storing the seed: the data chunks are first encoded using a systematic high-rate repeat-accumulate code, which produces redundant packets, and then converted to a sequence of oligonucleotides via a hybrid mapping scheme.
The mapping scheme ensures the absence of long homopolymers (but does not guarantee a balanced GC content), while the use of a specific repeat-accumulate code instead of a fountain code reduces the number of bits required for indexing and thus results in a higher overall information density.

In the first step, the authors segment the bit stream into 266-bit packets.
Redundant packets are generated using a rate-0.95 repeat-accumulate code (5\% of all generated packets are therefore redundant).
To each packet, the authors append 14 bits for indexing (as opposed to 32 bits dedicated to the storage of random seeds in~\cite{erlich2017dna}) and further 20 bits for the cyclic redundancy check (CRC), resulting in the overall packet length of 300 bits.
At the decoding stage, the packets that do not pass the CRC are simply discarded, which allows the authors to treat the DNA storage channel as an erasure channel, and the remaining packets are decoded using a message-passing decoder tailored to the erasure channel.
(This simplifies the decoder while missing an opportunity to reconstruct some of the corrupted packets using more sophisticated error-correction mechanisms instead of a CRC.)

In the second step, an attempt is made to convert the resulting 300-bit packets into sequences of nucleotides that conform to biological constraints.
After a straightforward binary-to-quaternary mapping from \{00,01,10,11\} to \{A,C,G,T\}, the resulting 150-letter sequence is permuted with one of the four predefined interleavers and then checked for compliance with the homopolymer-run constraint.
If one of the interleavers does produce a valid sequence, that sequence is synthesised, with an additional nucleotide that specifies the used interleaver appended at the end.
Otherwise, if no interleaver produces a valid sequence, the authors use an alternative variable-length constrained (VLC) mapping.
The VLC mapping employs a finite-state machine to represent possible run-length-limited sequences and uses the input data to drive the state transitions of that finite-state machine.
An extra nucleotide is added to distinguish between interleaver- and VLC-based mapping.

The proposed hybrid mapping guarantees the absence of homopolymer runs of length 4.
However, it does not guarantee a balanced GC content: using an interleaver does not change the GC content, so the GC-imbalanced sequences are mapped via VLC mapping straight away.
VLC mapping does not consider GC balance constraint explicitly and is not guaranteed to produce a sequence that satisfies it.
This approach achieves a higher overall information density than the fountain-code approach in~\cite{erlich2017dna}: 1.32 against 1.19 b/nt, respectively.

\subsection{Improved Read/Write Cost Tradeoff in DNA-Based Data Storage Using LDPC Codes~\texorpdfstring{\cite{chan2019improved}}{}}\label{sse:chan}
Most encoding schemes separate error correction in two stages: within- and across-strand coding.
Within-strand (or ``inner'') coding attempts to correct errors in individual sequence readouts and often uses a Reed-Solomon code.
Across-strand (or ``outer'') codes, in contrast, compensate for the loss of entire sequences and are therefore often designed for an erasure channel (e.g., the work~\cite{wang2019high} discussed in Section~\ref{sse:wang}).

This separation simplifies design but entails a loss in error-correcting performance.
In contrast, Chandak et al.~\cite{chan2019improved} use a single large regular LDPC code for both within- and across-strand error correction.
They use a block length of 256 thousand bits and split the codewords into 162-bit chunks that become 81-nt payloads of 150-nt strands to be synthesised.
Apart from the payload, the strands include two 25-nt PCR primers (which are used for amplification and can also be used for random access), 16-nt BCH-protected indices and a 3-nt sequence inserted in the middle of the payload to facilitate alignment.
The resulting information density depends on the rate of the LDPC code; the authors test their approach for a code rate of 0.91, 0.77 and 0.67 (which correspond to lengthening the bit stream by 10\%, 30\% and 50\%, respectively).
Using a rate of 0.91 yields a payload-only information density of 1.82 b/nt and an overall information density of 1.02 b/nt.
The input bits are permuted and encrypted to reduce the likelihood of violating biological constraints; a standard two-bit mapping is used afterwards.
This does not guarantee the absence of long homopolymers or GC imbalance.

During decoding, the reads are grouped by index and aligned.
A key idea of the work is to model the output of the alignment algorithm as an input-symmetric channel with binary input and ``vote counts'' as output.
Specifically, for each input bit, the channel produces a pair $(k_0, k_1)$, where $k_0$ and $k_1$ are the numbers of sequences that have $0$ and $1$ in the corresponding position after alignment, respectively.
The sum $k_0 + k_1$ is then the total number of used reads of the sequence, which the authors model as a Poisson-distributed random variable.
The disagreements between the strands are modelled to result from transmitting each of the $k_0 + k_1$ sequences over a binary symmetric channel with a predefined substitution probability $\epsilon$ ($\epsilon$ is set to $0.04$ in the experiments).
Under these assumptions, the authors derive a formula for channel log-likelihood ratios and use them as input to belief propagation decoding of the LDPC code.

Another key insight in~\cite{chan2019improved} is the elucidation of the trade-off between the cost of writing and reading in DNA storage: the more redundancy in the data, the higher the cost of synthesising the strands, but the fewer copies of the strands need to be sequenced for successful recovery.
In other words, the more expensive the writing of the data, the less expensive the reading will be.
The simplified model of the DNA storage channel the authors use to derive the log-likelihood ratios clearly reveals this trade-off.
They demonstrate that two approaches may have a similar writing cost but differ substantially in their reading cost, both measured in nucleotides per information bit.
Writing cost alone is therefore insufficient as a metric of the efficiency of an encoding scheme---reading cost must be considered as well.
(Reading cost can be obtained by dividing minimum coverage by the overall information density; we report both figures in Table~\ref{table:comparison}.)
By varying the LDPC code rate in their scheme, the authors obtain a simulated reading-vs-writing-cost curve and a set of points for five \textit{in-vitro} experiments.
Although differences in experimental setups complicate direct comparison, the authors demonstrate that their scheme can offer a better trade-off than those achieved in~\cite{erlich2017dna} (Section~\ref{sse:erlich}) and~\cite{Organick2018RandomStorage} (Section~\ref{sse:organick}). We remark that the length of the primers is not accounted for in their calculations.

\subsection{HEDGES Error-Correcting Code for DNA Storage Corrects Indels and Allows Sequence Constraints~\texorpdfstring{\cite{press2020hedges}}{}}\label{sse:hedges}
Previous encoding schemes we have discussed only provide within-strand error correction that is able to recover from substitution and deletion but not insertion events. Recent work by Press et al.~\cite{press2020hedges} recognises that for DNA storage to be viable, encoding schemes must be able to correct all three types of errors, i.e., substitutions, insertions and deletions, from a single read, and that the decoding error rate should be very low for a scheme to be scalable to large amounts of data. To this end, they have devised HEDGES (Hash Encoded, Decoded by Greedy Exhaustive Search), an error-correcting code that is able to repair all three basic types of DNA errors: insertions, deletions and substitutions.

\begin{figure}[ht]
    \includegraphics[width=0.96\linewidth]{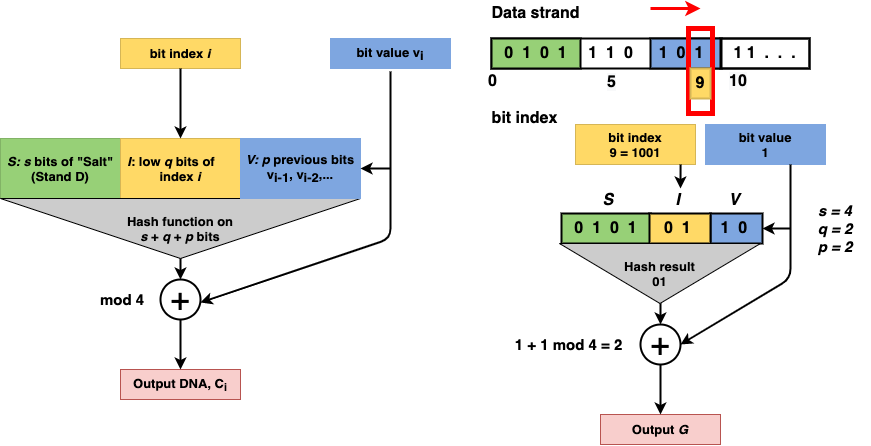}
    \caption{HEDGES encoding scheme. Each bit is encoded using a hash which is a function of \textit{s} bits of salt, the low \textit{q} bits of the index and \textit{p} previously encoded bits.}
    \label{fig:HEDGES-Encoding}
\end{figure}

\begin{figure}[ht]
    \includegraphics[width=0.96\linewidth]{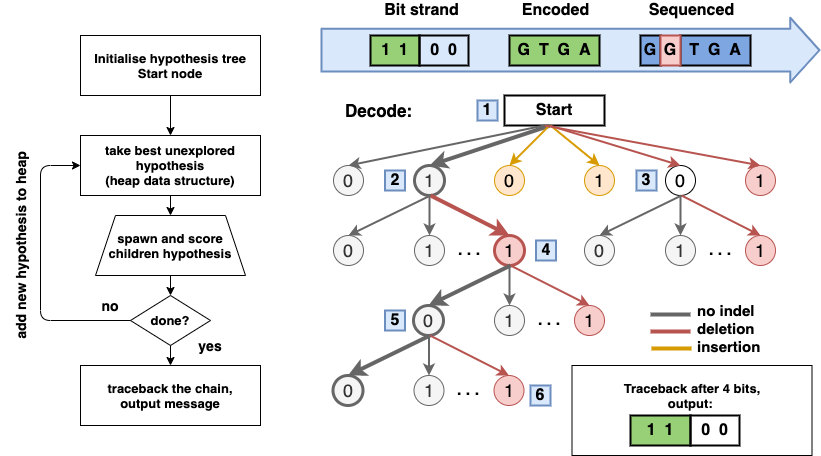}
     \caption{The HEDGES decoding algorithm. A greedy search is performed on an expanding tree of hypotheses. The hypothesis simultaneously guesses one or more message bits \textit{vi} , its bit position index \textit{i} and its corresponding DNA character position index \textit{k}. To limit exponential tree growth, a "greediness parameter" is used (see Supplementary Text of \cite{press2020hedges}).}
    \label{fig:HEDGES-Decoding}
\end{figure}

HEDGES is used as an inner code in combination with a Reed-Solomon outer code. Data is first split into blocks called message packets or DNA packets. An individual packet comprises an ordered set of 255 sequences, each of which has the same fixed length, allowing some insertions and deletions to be identified, and contains a packet ID and sequence ID. The bit stream is encoded with the HEDGES encoding algorithm, and the generated indexes for the sequence (packet ID and sequence ID) are protected using an encoding \textit{salt}. HEDGES encoding hashes the salt, the bit index and a number of previous bits and adds this to the value of the current bit modulo 4---the process is depicted in Figure~\ref{fig:HEDGES-Encoding}. A Reed-Solomon outer code is then applied diagonally across the DNA sequences within a message packet. Application of the salt enables errors to be converted to erasures that can later be corrected using the Reed-Solomon outer code.
HEDGES decoding uses an expanding tree of hypotheses for each bit, on which a greedy search is performed as shown in Figure~\ref{fig:HEDGES-Decoding}.

The authors suggest that the advantages of using HEDGES as an inner code are as follows: (i) sequences are of a fixed length (when uncorrupted), (ii) recovering synchronisation has the highest priority, (iii) deletions that are known are less problematic than substitutions that are unknown since an RS code can correct twice as many deletions as substitutions, (iv) burst errors within a single byte become less problematic than distributed errors as the used RS code corrects errors at the level of bytes, and (v) error-free messages can be yielded despite residual errors so long as the byte-level errors and deletions are within the RS code's error-correcting capability.

To test the HEDGES algorithm, they used both \textit{in-silico} simulation and \textit{in-vitro} synthesis of pools of oligonucleotides. For \textit{in-silico} testing, nucleotides were artificially generated across binned ranges of code rates and error probabilities. For each code rate $r$ in (0.166, 0.250, 0.333, 0.500, 0.600, 0.750) and each estimated error probability $P_{\mathsf{err}}$ in (0.01, 0.02, 0.03, 0.05, 0.07, 0.10, 0.15), 10 packets ($\sim 10^6$ nucleotides) were encoded. The error rate for bit and byte errors was computed for the HEDGES output for each pair $(r, P_{\mathsf{err}})$. This was then used to extrapolate to the Petabyte and Exabyte scales. They observe that the rate of byte errors was less that $8\times$ that of the rate expected for independent bit errors and postulate that this is because decoding errors tend to occur in bursts.

For their \textit{in-vitro} work, 5,865 DNA sequences of length 300 were synthesised, each with a payload of 256 nucleotides flanked by two primers of 23 nucleotides at their 5' (head) and 3' (tail) ends. These were degraded at high temperature and then sequenced with a coverage of approximately 50 (the authors perform a sub-sampling experiment and demonstrate that a coverage of about 5 is in fact sufficient to recover the data). They calculated the end-to-end DNA error rates---i.e., errors that are introduced during synthesis, sample handling and storage, preparation, and finally sequencing---for insertions, deletions and substitutions, which were observed to be to be 17\%, 40\% and 43\%, respectively.
The authors used a mixture of HEDGES code rates in their experiment; using the HEDGES rate of 0.75 yields an overall information density of 1 b/nt and a payload-only information density of 1.3 b/nt.

\section{Discussion}
Table~\ref{table:comparison} compares the previously discussed approaches with respect to their storage mechanism (organism or microplate), information density (b/nt), coverage, error detection and correction mechanisms as well as the consideration of biological constraints (whether the approach avoids homopolymers or balances the GC content). We present both the overall and the payload-only information density as discussed in Section~\ref{sse:density}.


With respect to the storage medium, early approaches tend to store information in organisms, whereas later ones store the synthetic DNA on microplates. Storing information in living organisms has the benefit that it can be passed on in highly resistant organisms, e.g., bacteria resistant to adverse conditions (extreme temperatures and similar), over many generations and thus for a very long time. At the same time, however, DNA recombination during cell division is an error-prone process that may introduce substantial errors in the stored information. Using living organisms also limits the amount of information that can be stored because the sequence inserted into the organism cannot be too long without incapacitating the organism. 

Given the dependence of operations on DNA on biochemical and molecular processes, we have also looked at whether the approaches we reviewed adhere to two biological constraints, i.e., whether long homopolymers and extremes in the GC content are avoided. Both of these biological constraints are necessary to improve the stability of the sequence and reduce errors during synthesis and sequencing---more than half of the reviewed approaches do not adhere to both of these constraints, of which several adhere to either preventing homopolymer runs or avoiding GC extremes (but not both). Only a few approaches adhere to both constraints.

Inherent in the use of DNA as an information storage medium are unwanted substitution, deletion and insertion events. These errors can occur during both synthesis and sequencing, both \textit{in vivo} and, notwithstanding the general stability of DNA, \textit{in vitro}, e.g., due to temperature, UV radiation, pH, etc. Whilst living organisms have molecular mechanisms in place to repair these errors to some extent, the use of DNA as a storage medium for arbitrary data will require the application of algorithms and design approaches to mitigate errors. Furthermore, the error rate will determine the extent to which DNA storage approaches can scale. Hence, we have also analysed the error detection and correction techniques used in each approach. The error detection and correction techniques span from none to replication, where the DNA sequences are replicated and errors are corrected through a majority vote (multiple copies of sequences can be pre-synthesised to correct all errors or post-amplified after synthesis to correct sequencing errors), to Reed-Solomon and other error-correcting codes.

The maximum theoretical information density of DNA is 455 EB/g, allowing vast amounts of information to be stored in this medium. However, the state of current DNA synthesis technologies means the process is expensive. Consequently, the information density achieved in the experiments of each approach, as well as the data stored, is important, and has been reviewed and compared. Although microdot project~\cite{microdot} (Section~\ref{sse:microdot}) and ``Blighted by Kenning,'' the artwork DNA storage project by Jarvis~\cite{kenning} (Section~\ref{sse:kenning}), have the highest information density achievable without compression (each have managed 2 b/nt), both approaches neither adhere to any biological constraints nor use any error detection or correction. They will, thus, not be scalable and would be impractical for application to arbitrary data storage.

Some approaches attempted a sub-sampling experiment to determine the minimum coverage sufficient for successful decoding.
Despite the differences in experimental setups, Table~\ref{table:comparison} shows that a higher information density tends to require a higher coverage, with a coverage of around 10 being achievable for quite high information densities. A coverage of around 5, on the other hand, requires lowering the information density significantly.

Work done by Press et al.~\cite{press2020hedges} (Section~\ref{sse:hedges}) uses a combination of their HEDGES encoding as an inner code and a Reed-Solomon outer code. It is the first approach that corrects all types of errors (insertions, deletions and substitutions) within individual strands and achieves error-free petabyte-scale storage while tolerating up to 10\% of the nucleotides being erroneous.

\fontsize{7}{7}\selectfont

\renewcommand{\arraystretch}{1.5}

\begin{longtable}{ p{2.2cm}  p{1.2cm}  p{1.8cm}  p{1.3cm}  p{1.3cm}  p{1cm} p{1.7cm}p{1.6cm}  p{1.3cm} }
    
    \caption{Comparison of DNA encoding schemes. An asterisk indicates that the information density figure includes compression; a dagger indicates that the authors did not attempt a sub-sampling experiment to stress-test their approach under reduced coverage.}
    \label{table:comparison}\\
    
    \hline
    \textbf{Approach} \newline \textbf{(Reference)} & \textbf{Storage} \newline \textbf{Mechanism} & \textbf{Encoding} \newline \textbf{Alphabet}  &
    \textbf{Overall} \newline \textbf{Information} \newline \textbf{Density (b/nt)} & \textbf{Payload} \newline \textbf{Information} \newline \textbf{Density (b/nt)} & \textbf{Coverage} & \textbf{Error} \newline \textbf{Correction} \newline \textbf{and} \newline \textbf{Detection} & \textbf{Biological} \newline \textbf{Constraints} & \textbf{Access} \newline \textbf{ Mechanism} \tabularnewline
    \hline
    \endfirsthead
    
    \hline
    \textbf{Approach} \newline \textbf{(Reference)} & \textbf{Storage} \newline \textbf{Mechanism} & \textbf{Encoding} \newline \textbf{Alphabet}  &
    \textbf{Overall} \newline \textbf{Information} \newline \textbf{Density (b/nt)} & \textbf{Payload} \newline \textbf{Information} \newline \textbf{Density (b/nt)} & \textbf{Coverage} & \textbf{Error} \newline \textbf{Correction} \newline \textbf{and} \newline \textbf{Detection} & \textbf{Biological} \newline \textbf{Constraints} & \textbf{Access} \newline \textbf{ Mechanism} \tabularnewline
    \hline
    \endhead
    
    \hline
    \endlastfoot
    
    Microvenus~\cite{Davis1996Microvenus} \newline Section~\ref{sse:microvenus}  & Organism  & {[0,1] (bit stream)} & 1.25* & 1.94* & --- & --- & --- & sequential \\ 
    
    Genesis~\cite{Kac1999GENESIS} \newline Section~\ref{sse:genesis} & Organism & [A--Z] & 1.52* & 1.52* & --- & --- & --- & sequential \tabularnewline
    
    Bancroft et al.~\cite{Bancroft2001Long-TermDNA} \newline Section~\ref{sse:bancroft} & Microplate & [A--Z] & 0.94 & 1.65 & --- & --- & --- & random \\

    Wong et al.~\cite{Wong2003OrganicApproach} \newline Section~\ref{sse:wong} & Organism & [A--Z, a--z, 0--9, !] & 1.54 & 1.62 & --- & --- & --- & sequential \\ 
    
    Microdot~\cite{microdot} \newline Section~\ref{sse:microdot} & Microplate & [A--Z, a--z, !] & 1.27 & 2 & --- & --- & --- & sequential \\ 

    Huffman codes~\cite{Smith2003SomeDNA} \newline Section~\ref{sss:smith_huffman} & Microplate & [A--Z] & 2.27* & 2.27* & --- & --- & --- & sequential \\

    Comma codes~\cite{Smith2003SomeDNA} \newline Section~\ref{sss:smith_comma} & Microplate & [A--Z, a--z, 0-9, !] & 1.17 & 1.17 & --- & detection & balanced GC content & sequential  \\

    Alternating codes~\cite{Smith2003SomeDNA} \newline Section~\ref{sss:smith_alternating} & Microplate & [A--Z, a--z, !] & 1 & 1 & --- & detection & no homopolymers & sequential  \\

    Yachie et al.~\cite{Yachie2007Alignment-basedOrganisms} \newline Section~\ref{sse:yachie} & Organism & [A--Z, a--z, 0--9, !] & 0.49 & 0.5 & --- & replication (pre) & --- & sequential \\
    
    Portney et al.~\cite{portney} \newline Section~\ref{sse:portney} & Organism & [A,E,I,M,O,S,U,Z] & 0.11 & 0.2 & --- & replication (pre) & --- & random \\

    Ailenberg et al.~\cite{Ailenberg2009AnDNA} \newline Section~\ref{sse:ailenberg} & Organism & [A--Z, a--z, 0--9, !] & 1.9* & 2* & --- & --- & --- & sequential\\

    Kenning~\cite{kenning} \newline Section~\ref{sse:kenning} & Organism & [A--Z]  & 2 & 2 & --- & --- & --- & sequential\\

    Church et al.~\cite{Church2012Next-generationDNA} \newline Section~\ref{sse:church} & Microplate & [0,1] (bit stream) & 0.6 & 1 & 3000\textsuperscript{\textdagger} & replication (post) & no homopolymers; balanced GC content & random \\
 
    Goldman et al.~\cite{Goldman2013TowardsDNA} \newline Section~\ref{sse:goldman} & Microplate & [0,1] (bit stream) & 0.22* & 0.39* & 10 & detection and correction & no homopolymers & sequential \\
    Grass et al.~\cite{Grass2015RobustCodes} \newline Section~\ref{sse:grass} & Microplate & [0,1] (bit stream) & 0.86 & 1.25 & 372\textsuperscript{\textdagger} & detection and correction & no homopolymers; balanced GC content & sequential \\
    Bornholt et al.~\cite{Bornholt2016ASystem} \newline Section~\ref{sse:bornholt} & Microplate & [0,1] (bit stream) & 0.59* & 1.05* & 40 & replication (pre) & no homopolymers & random \\

    Yazdi et al.~\cite{HosseinTabatabaeiYazdi2017PortableStorage} \newline Section~\ref{sse:yazdi} & Microplate & [0,1] (bit stream) & 1.72 & 1.75 & 200\textsuperscript{\textdagger} & replication (post) & no homopolymers; balanced GC content & random\\

    Oligoarchive DB~\cite{oligoarchive} \newline Section~\ref{sse:oligoarchive} & Microplate & [0,1] (bit stream) & 1.79* & 2.73* & $2 \cdot 10^5$\textsuperscript{\textdagger} & detection & no homopolymers & random\\
    Oligoarchive Join~\cite{oligoarchive} \newline Section~\ref{sse:oligoarchive} & Microplate & [0,1] (bit stream) & 1.43 & 1.43 & --- & detection and correction & --- & random \\
     
    JPEG~\cite{dimopoulou2020image} \newline Section~\ref{sse:jpeg} & Microplate & [0,1] (JPEG files) & --- & 1.6* & --- & detection & no homopolymers; balanced GC content & sequential \\

    200MB storage~\cite{Organick2018RandomStorage} \newline Section~\ref{sse:organick} & Microplate & [0,1] (bit stream) & 0.81 & 1.3 & 5 & detection and correction & no homopolymers & random \\
    Fountain codes~\cite{erlich2017dna} \newline Section~\ref{sse:erlich} & Microplate & [0,1] (bit stream) & 1.19 & 1.75 & 10.4 & detection and correction & no homopolymers; balanced GC content & random \\

    RA codes~\cite{wang2019high} \newline Section~\ref{sse:wang} & Microplate & [0,1] (bit stream) & 1.32 & 1.75 & 10 & detection and correction & no homopolymers & random \\

    Large LDPC~\cite{chan2019improved} \newline Section~\ref{sse:chan} & Microplate & [0,1] (bit stream) & 1.02 & 1.82 & 12.4 & detection and correction & --- & random \\

    HEDGES~\cite{press2020hedges} \newline Section~\ref{sse:hedges} & Microplate & [0,1] (bit stream) & 1 & 1.3 & 5.2 & detection and correction & no homopolymers; balanced GC content & sequential \\
\end{longtable}

  \normalsize

Finally, we have reviewed the data-retrieval method, i.e., sequential or random access, that each DNA storage approach implements. Accessing only part of information stored in DNA using sequential access requires sequencing to reconstruct all stored information, which is slow and costly. This method is used by most of the reviewed DNA storage approaches, particularly the earlier ones. On the other hand, random access allows subsets of the data to be selectively retrieved by amplifying only select regions of the DNA using primers associated with the regions of interest. This is cheaper, more efficient and provides faster access. Several approaches use random access, including the early work by Bancroft et al.~\cite{Bancroft2001Long-TermDNA} (Section~\ref{sse:bancroft}), a similar method by Bornholt et al.~\cite{Bornholt2016ASystem} (Section~\ref{sse:bornholt}), an in-depth design by Yazdi et al.~\cite{HosseinTabatabaeiYazdi2017PortableStorage} (Section~\ref{sse:yazdi}) and both the Oligoarchive DB and join projects~\cite{oligoarchive} (Section~\ref{sse:oligoarchive}).

\section{Conclusions \& Outlook}
Encoding approaches for DNA storage have evolved drastically over time. Initially primarily used for artistic purposes and understood as proof of principle, the encodings and their features have evolved considerably, with the latest approaches providing a solid basis for viable, long-term DNA data storage. Fundamental biological constraints on long homopolymers and extreme GC content have generally been considered in encodings early on. Also, important features such as error detection and correction have been incorporated in most encodings, particularly in the later ones. More advanced features, such as random access through the incorporation of primers (to allow for selective reading of regions of interest, thereby avoiding the need to sequence all of the information), are rather new and are only used where required.

Albeit a comparison of information density (bits/nucleotide)---a crucial metric given today's synthesis cost---is difficult given the diversity of features supported by different encodings, it has dropped over the years as more constraints have been considered and more features incorporated.
The encodings (and their features) appear to converge, with the vast majority of recent encodings considering biological constraints and incorporating some form of error detection and correction.

Future encodings are likely to consider additional constraints (e.g., storage, subsequences unsuitable for synthesis or sequencing, etc.) to improve stability in storage or to lower errors in synthesis and sequencing. Further, advances in synthesis and sequencing technology will change their error characteristics, which will affect the requirements for error correction and detection codes. The direction of this evolution is difficult to foretell; on the one hand, synthesis and sequencing could become more accurate, reducing the need for error correction. On the other hand, the push towards higher throughput and lower cost of synthesis and sequencing may lower their accuracy instead, bringing error correction to the fore of DNA storage research. Advances in encoding are also likely in the application of different or new error correction codes (e.g., polar codes~\cite{polarcodes}) which may offer better protection against insertion, deletion and substitution errors in the DNA storage channel. A significant improvement in information density, however, is unlikely.


\bibliographystyle{ACM-Reference-Format}
\bibliography{references}

\appendix



%
%

\end{document}